\documentclass[authoryear,preprint,12pt]{elsarticle}

\usepackage{amssymb}
\usepackage{amsthm}
\usepackage{amsmath}

\usepackage[left]{lineno} % default option is 'left'
\makeatletter
\def\makeLineNumberLeft{%
  \linenumberfont\llap{\hb@xt@\linenumberwidth{\LineNumber\hss}\hskip\linenumbersep}% left line number
  \hskip\columnwidth% skip over column of text
  \rlap{\hskip\linenumbersep\hb@xt@\linenumberwidth{\hss\LineNumber}}\hss}% right line number
\leftlinenumbers% Re-issue [left] option
\makeatother

\usepackage[dvipsnames]{xcolor}
\usepackage{enumitem}

\usepackage{caption}
\usepackage{subcaption}
\usepackage{tabularx}
\usepackage{booktabs}

\usepackage[acronym]{glossaries}
\newacronym{aic}{AIC}{Akaike information criterion}
\newacronym{bm}{BM}{block maxima}
\newacronym{cdf}{CDF}{cumulative distribution function}
\newacronym{cv}{CV}{cross-validation}
\newacronym{dcc}{DCC}{dynamic conditional correlation}
\newacronym{emr}{EMR}{expected maximum size of risk events}
\newacronym{es}{ES}{Expected Shortfall}
\newacronym{garch}{GARCH}{generalized autoregressive conditional heteroskedasticity}
\newacronym{iid}{i.i.d.}{independent and identically distributed}
\newacronym{lasso}{LASSO}{least absolute shrinkage and selection operator}
\newacronym{mab}{MAB}{multi-armed bandit}
\newacronym{mle}{MLE}{maximum likelihood estimation}
\newacronym{mpmr}{MPMR}{most probable maximum size of risk events}
\newacronym{mse}{MSE}{mean squared error}
\newacronym{mvo}{MVO}{Mean-variance optimization}
\newacronym{pd}{p.d.}{positive definite}
\newacronym{ppf}{PPF}{percent point function}
\newacronym{rv}{RV}{random variable}
\newacronym{se}{SE}{scaling exponent}
\newacronym{soc}{SOC}{self-organized criticality}
\newacronym{spt}{SPT}{stochastic portfolio theory}
\newacronym{tdc}{TDC}{tail dependence coefficient}
\newacronym{ti}{TI}{tail index}
\newacronym{var}{VaR}{Value-at-Risk}
\newacronym{wmle}{WMLE}{weighted maximum likelihood estimation}

\usepackage{hyperref} \hypersetup{colorlinks=true, linkcolor=blue, filecolor=magenta, urlcolor=cyan, pdftitle={Overleaf Example}, pdfpagemode=FullScreen,}

\usepackage{cleveref}
\usepackage{natbib}
\let\today\relax %% hide date
\makeatletter
\def\ps@pprintTitle{%
    \let\@oddhead\@empty
    \let\@evenhead\@empty
    \def\@oddfoot{\footnotesize\itshape
         {Submitted preprint} \today \hfill}%
    \let\@evenfoot\@oddfoot
    }
\makeatother

\begin{document}
%%\linenumbers

\begin{frontmatter}



\title{
A General Framework for Portfolio Construction Based on Generative Models of Asset Returns
}

\author[rmi]{Tuoyuan Cheng\corref{cor1}}
 % \ead{tuoyuan.cheng@nus.edu.sg}
\author[rmi,math]{Kan Chen}
 % \ead{kan.chen@nus.edu.sg}

 \cortext[cor1]{Corresponding author: tuoyuan.cheng@nus.edu.sg}

 \address[rmi]{Risk Management Institute, National University of Singapore, 04-03 Heng Mui Keng Terrace, I3 Building, Singapore, 119613, Singapore}
 \address[math]{Department of Mathematics, National University of Singapore, Level 4, Block S17, 10 Lower Kent Ridge Road, Singapore, 119076, Singapore}

\begin{abstract}
%\begin{linenumbers}
In this paper, we present an integrated approach to portfolio construction and optimization, leveraging high-performance computing capabilities. We first explore diverse pairings of generative model forecasts and objective functions used for portfolio optimization, which are evaluated using performance-attribution models based on \gls{lasso}. We illustrate our approach using extensive simulations of crypto-currency portfolios, and we show that the portfolios constructed using the vine-copula generative model and the Sharpe-ratio objective function consistently outperform. To accommodate a wide array of investment strategies, we further investigate portfolio blending and propose a general framework for evaluating and combining investment strategies. We employ an extension of the multi-armed bandit framework and use value models and policy models to construct eclectic blended portfolios based on past performance. We consider similarity and optimality measures for value models and employ probability-matching (``blending") and a greedy algorithm (``switching") for policy models. The eclectic portfolios are also evaluated using \gls{lasso} models. We show that the value model utilizing cosine similarity and logit optimality consistently delivers robust superior performances. The extent of outperformance by eclectic portfolios over their benchmarks significantly surpasses that achieved by individual generative model-based portfolios over their respective benchmarks.  
%\end{linenumbers}
\end{abstract}


%\item 


\end{frontmatter}

% \linenumbers

%% reset glossaries to start from explanations again then abbreviations.
\glsresetall

\section{Introduction}
\label{sec:intro}
\noindent
Since the seminal works of Markowitz, Finetti, and Roy seven decades ago, significant efforts from academia and industry have been devoted to portfolio theories \citep{markowitz_portfolio_1952,markowitz_early_1999,markowitz_finetti_2006,roy_safety_1952,rubinstein_markowitzs_2002}.
The \gls{mvo} leverages the first and second moments of the portfolio return to establish a risk-return trade-off and define the efficient frontier. \gls{mvo} operates without assumptions of Gaussian-distributed returns or quadratic investor utility functions \citep{kolm_60_2014,markowitz_portfolio_2010}.
Various paradigms rooted in classical and Bayesian statistical decision theory, as well as their causal extensions, have been applied to address different utility configurations. These encompass von Neumann and Morgenstern's theory, Savage's theory, rank-dependent utility theory, and prospect theory \citep{friedman_utility_1948, heckerman_decision-theoretic_1995,howard_decision_1972,kahneman_prospect_1979,lewis_causal_1981,pearl_causal_1995,ramsey_foundations_1931,savage_theory_1951,schmidt_alternatives_2004,von_neumann_theory_1944}. 
Beyond those foundational theories, the Black-Litterman model, a Bayesian approach for market-based shrinkage in portfolio optimization, incorporates investor views to refine market equilibrium implied expected returns and to enhance asset return covariance matrix estimation \citep{kolm_blacklitterman_2021}. 
The risk parity approach distributes portfolio risk evenly across covered assets, aiming to withstand market downturns \citep{qian_financial_2005}. 
The \gls{spt} employs Brownian motion to construct stochastic processes for asset prices, offering a descriptive instead of normative approach
\citep{fernholz_diversity_1999, bensoussan_stochastic_2009}. 
Higher-moment portfolio theory recognizes the non-Gaussian properties of asset returns, aiming for improved returns and curtailed tail risks \citep{malevergne_higher-moment_2005}.
Robust portfolio theory acknowledges input estimate uncertainties, accounting for parameter uncertainty and model misspecification \citep{xidonas_robust_2020}.
In summary, a rich tapestry of compelling portfolio theories, each accompanied by corresponding objective functions, is documented in the literature, offering a diverse array of options for portfolio optimization.
\\\\
The prevailing \gls{mvo} framework integrates risk and alpha models. While risk models assess portfolio return variance, alpha models predict idiosyncratic expected returns. 
The advent of enhanced computing capabilities has ushered in a new era for portfolio optimization, driven by generative models capable of simulating multivariate asset returns across a spectrum of market scenarios \citep{markowitz_meanvariance_2014,kolm_60_2014}.
Generative model simulations drawn from model-implied predictions serve multiple crucial purposes, including model validation, goodness-of-fit assessment, software implementation evaluation, embedded research hypotheses refinement, and forecasts visualization \citep{mcelreath_statistical_2020,weinzierl_pillars_2022}.
By harnessing generative model simulations for multivariate asset returns, we gain the capability to optimize performance metrics targeting moments, quantiles, and ordinals of the univariate portfolio return distribution. 
\\\\
Amidst the multitude of portfolio theories and the availability of generative models, decision-making becomes more diverse and fragmented. Individual investors often resort to a diverse set of mental models in their Swiss Army knife toolboxes \citep{munger_poor_2008}. Discretionary portfolio managers working in silos allocate capital and influence institutional holdings. Research-oriented investment firms merge signals, blend portfolios, and make strategic shifts at a meta-level \citep{de_prado_10_2018}. Market participants must navigate a complex landscape of risk-return trade-offs, time horizons, tax considerations, liquidity needs, and mandates. It might be advantageous for investors to opt for an eclectic portfolio, which can accommodate a range of portfolio theories and strategies, instead of adhering strictly to a single combination of proxy objective functions and generative models.
\\\\
The literature has seen noteworthy efforts in addressing eclectic portfolio construction via portfolio blending through the lens of \gls{mab} problem-solving. \cite{shen_portfolio_2016} introduced an approach that leverages Thompson sampling for online portfolios, where they assign blending ratios $\psi$ as probabilities for selecting from three basis portfolios, and adopt Bayesian decision rules to update the corresponding distribution functions. Assuming Bernoulli distribution for portfolio performances, the algorithm sequentially determines $\psi$ to encapsulate a spectrum of investment criteria and market views. Remarkably, it demonstrates a distinct advantage over any individual basis portfolio across various market scenarios. In a similar vein, \cite{fujishima2022multiple} experiment with five basis portfolios within the \gls{mab} framework and assume Dirichlet distribution for portfolio performances to determine the blending ratio $\psi$. Their backtests show the potential benefits of blending in a cost-free transaction environment. Within the \gls{mab} setting, the agent maintains a value model to evaluate the optimality among arms, and a policy model to decide which arm to pull either deterministically or stochastically. In defining the value model, it is imperative to establish a portfolio performance measure that not only accentuates decision quality but also remains comparable across a diverse array of portfolios and time steps. When crafting the policy model, achieving a balance in the concentration-diversification dilemma at the portfolio level becomes a necessity.
\\\\
Cryptocurrencies have notably high correlations and their covariance matrices often possess elevated condition numbers. This attribute necessitates the application of advanced techniques in multivariate dependence modeling, asset returns generative modeling, and portfolio construction. This demand underscores the importance of evaluating and optimizing portfolio construction methodologies. Moreover, global cryptocurrency exchanges operate $24/7$, facilitating the trading of homogeneous products with limited regulatory oversight. A multitude of participants have access to real-time price information, entering and exiting positions with the goal of profit maximization. These markets, with their unique characteristics, are still emerging and under-researched. This emerging terrain presents a fresh opportunity for advancing portfolio theory, which holds the potential to not only improve the management of cryptocurrency portfolios but also yield insights with broader applicability across financial markets.
\\\\
In this paper, we embark on a comprehensive exploration of our portfolio construction framework, examining its foundation elements and efficacy evaluation.  We first investigate and evaluate diverse pairings of generative models and objective functions. We then explore portfolio blending for constructing eclectic portfolios aligned with a wide array of investment strategies. We adapt the multi-armed bandit environment to the domain of portfolio blending. Within this \gls{mab} context, we propose similarity and optimality measures for value function estimates, and employ probability-matching (termed ``blending") or greedy algorithm (termed ``switching") for action selection policies. We show a framework for constructing superior eclectic portfolios through a thorough performance analysis of value model and policy model combinations. By conducting this exploration within the dynamic landscape of cryptocurrencies, we delve into an emerging area of finance that draws inspiration from multiple existing portfolio theories and practices. By doing this, we aim to provide solutions and perspectives that have relevance not only in the cryptocurrency domain but also hold applicability across wider financial markets.

\section{Constructing generative model-based portfolio: three essential components}
\label{sec:genport}
\noindent
The multi-period portfolio construction objective function can be formulated as follows:

\begin{align}
    \underset
    {\mathbf{w}_{0,1}, \hdots, \mathbf{w}_{0,t}, \dots, \mathbf{w}_{0,T}}
    {\max}~
    &{\prod_{t=1}^{T}~
    \bigg(
    1+r_{p,t}
    \left(\mathbf{w}_{0,t},\mathbf{r}_{t}
    \big|
    \mathbf{w}_{1,t-1},c\right)
    \bigg)
    }
    \nonumber
    \\=
    \underset
    {\mathbf{w}_{0,1}, \hdots, \mathbf{w}_{0,t}, \dots, \mathbf{w}_{0,T}}
    {\max}~
    &{\prod_{t=1}^{T}~
    \bigg(
    1+\mathbf{w}_{0,t}\mathbf{r}_{t}^T
    - 
    c \left\|\mathbf{w}_{0,t}-\mathbf{w}_{1,t-1}\right\|_1\bigg)
    } 
    \label{eq:objfun_ultimate}
    \\s.t.~ 
    &\text{constraints on }{\mathbf{w}_{0,1}, \dots, \mathbf{w}_{0,t}, \dots, 
    \nonumber
    \mathbf{w}_{0,T}}
\end{align}

\noindent
where $\mathbf{r}_t\in \mathbb{R}^{1\times D}$ is the row vector of asset returns from day $t-1$ to $t\in\{1,2,\dots, T\}$, $D$ is the number of assets, T denotes the investment horizon, ${\mathbf{w}_{1,t-1}}\in\mathbb{R}^{1\times D}$ is the row vector of asset weights in the portfolio near the close of day $t-1$, and $c$ denotes the percentage transaction cost associated with weight changes. The scalar return of the portfolio $r_{p,t}$ on day $t$, depends on $\mathbf{r}_t$ and $\mathbf{w}_{0,t}$, as well as previous weights $\mathbf{w}_{1,t-1}$ and transaction costs $c$. The decision vector $\mathbf{w}_{0,t} \in \mathbb{R}^{1\times D}$ is the row vector of asset weights of the portfolio at the close of day $t-1$. 
\\\\
Note that $\mathbf{r}_t$ remains unknown when $\mathbf{w}_{0,t}$ is being determined. Therefore, all objective functions act as temporary approximations to \cref{eq:objfun_ultimate}, relying either explicitly or implicitly on forecasts of multivariate asset returns $\mathbf{r}_t$.
The cumulative product in \cref{eq:objfun_ultimate}, which equally weights early and late portfolio gains, caters to rational, patient, and long-term investors. In practical applications, researchers often focus on one-period models due to the complexities associated with forecasting multivariate asset returns under real-world constraints \citep{lezmi_multi-period_2022}. 
\\\\
In this paper, we optimize a one-period proxy objective function $f(\cdot)$ (\cref{eq:objfun}) targeting various performance metrics and risk-return appetites, subject to certain user-specific and time-specific constraints:

\begin{align}
    \underset
    {\mathbf{w}_{0}}
    {\max}~
    &{
    f(\mathbf{w}_0|R^\mathbb{P}, \mathbf{w}_{1}, c, v)
    }
    \label{eq:objfun}
    \\
    s.t.~ 
    &
    \|\mathbf{w}_0\|_1=1,
    ~
    w_{0,d}\in\left[-\frac{m}{D},\frac{m}{D}\right],
    ~
    d\in\{1,2,\dots,D\}
    \nonumber
\end{align}

\noindent
where the decision vector ${\mathbf{w}_0}\in \mathbb{R}^{1\times D}$ is the row vector of asset weights after rebalancing, ${\mathbf{w}_1} \in \mathbb{R}^{1\times D}$ is the row vector of asset weights before rebalancing. The matrix $R^\mathbb{P}\in \mathbb{R}^{N\times D}$ is filled with multivariate simulated asset returns drawn from generative models. Here $N$ is the number of samples, $c$ is the transaction cost, $v$ is the scalar transaction-cost aversion coefficient (it is used to quantify a hurdle for trading), and $m$ is the multiplier used in the weight box constraint.
\\\\
The framework encompasses three essential components: generative models, proxy objective functions, and optimization constraints. Optimization constraints are indispensable but case-specific in real-world applications. In an order that is easier to comprehend, we successively discuss constraints, proxy objective functions, and generative models.

\subsection{Constraints}
\noindent
Constraints play a pivotal role in enhancing out-of-sample performance by mitigating volatility, preventing over-concentration, and reducing downside risk. Constraints can originate from a variety of sources: regulatory policies, client guidelines, discretionary exposure limits, compliance requirements, and considerations related to market volume, reflexivity, and liquidity \citep{kolm_60_2014,soros_fallibility_2013}. Selecting and formulating constraints often involves a degree of subjectivity, making it more an eclectic art than a science, as it necessitates judgment and consideration of multiple factors. 
Markowitz emphasized that if the one-sum \citep{roy_safety_1952} is the sole constraint on portfolio choice, the resulting negative weights are far from an accurate representation of real-world positions \citep{markowitz_portfolio_2010}.
\\\\
In this study, without loss of generality, we set the following constraints when solving \cref{eq:objfun}:

\begin{align*}
    \|\mathbf{w}_0\|_1
    =&
    1
    \\
    w_{0,d}
    \in& 
    \left[-\frac{m}{D},\frac{m}{D}\right]
\end{align*}

\noindent
where $D$ is the number of assets covered in each portfolio, and we typically use $m=5$ for the weight constraint. This $L1$-norm equality constraint is conservative and allows for both equally long weights and equally short weights. In practice, investors can utilize $\|\mathbf{w}_0\|_1=l,~l>0$ to meet permitted leverages levels.

%%{\color{blue}(Only this L1 form constraint allows for the above cosine expression of parity portfolios in \cref{eq:eqlong,eq:eqshort,eq:stddevparity}.)}
% https://quant.stackexchange.com/questions/74475/l1-norm-equality-constraints-in-portfolio-optimization-pros-and-cons

\subsection{Proxy objective functions}
\subsubsection{Kelly portfolio}
\noindent
According to the standard Kelly criterion \citep{kelly_new_1956,thorp_portfolio_1975}, when allocating $1-w$ to a risk-free zero return asset and $w\in(0,1)$ to a risk-bearing asset which has a probability $p\in(0,1)$ to gain $b$ and $(1-p)$ to lose $a$, the objective function for determining $w$ (assuming the number of periods $N\rightarrow\infty$ and no transaction costs) is:

\begin{align*}
    \underset{w}{\max}~&
    Np\log(1+wb)+N(1-p)\log(1-wa)
    \\
    s.t.~& 
    w \in (0,1)
\end{align*}

\noindent
Let $\mathbf{w}_0=[w,1-w]$ be a row vector for weights, $\mathbf{p}=[p,1-p]^T$ be a column vector for probability mass function, and $R=\left(\begin{matrix}b&0\\-a&0\end{matrix}\right)$ be a matrix of asset return scenarios. To generalize Kelly's objective function for multiple assets, we use the Hadamard (element-wise) product notation $\odot$:

\begin{align}
    \max~&
    \mathbf{p}^T  \log(1 + R \mathbf{w}_0^T)
    \nonumber
    \\=
    \max~&\sum\sum
    \mathbf{p} \odot \log(1 + R\odot \mathbf{w}_0)
    \nonumber
    \\\approx
    \max~&\sum\sum
    \log(1 + \mathbf{p} \odot R \odot \mathbf{w}_0)
    \label{eq:kelly_approx}
\end{align}

\noindent
The approximation in \cref{eq:kelly_approx} holds when the absolute values of elements in $R\odot\mathbf{w}$ are much smaller than $1$, which is typically the case in well-diversified portfolios over a single period.
Let $R^\mathbb{P} \in \mathbb{R}^{N\times 2}$ be a tall matrix with $p$ proportion of rows as $[b,0]$ and the rest as $[-a,0]$. After absorbing the effect of $\mathbf{p}$ into $R^\mathbb{P}$, we have:

\begin{align*}
    \max~&\sum\sum \log(1+R^\mathbb{P}\odot\mathbf{w}_0)
    \\
    \approx
    \max~&\sum\sum (R^\mathbb{P}\odot\mathbf{w}_0-\frac{(R^\mathbb{P}\odot\mathbf{w}_0)^2}{2}+\frac{(R^\mathbb{P}\odot\mathbf{w}_0)^3}{3}-\frac{(R^\mathbb{P}\odot\mathbf{w}_0)^4}{4})
\end{align*}

\noindent
This proxy objective function can accommodate more assets with longer $\mathbf{w}_0$ and a wider $R^\mathbb{P}$.  
%%The entropy of the multi-asset returns distribution could affect the height of $R$. 
The Kelly portfolio employs portfolio log return $\log(1+r_p)$ instead of portfolio simple return $r_p$ as its objective function, leading to diminishing marginal utility and tail loss aversion. Unlike traditional portfolio construction, where risk-free assets with zero variances and covariances are considered separately as a leveraging concern, the Kelly portfolio can be seamlessly integrated into the asset portfolio construction process.
\\\\
In this paper, we use the following Kelly portfolio proxy objective function:

\begin{align*}
    \max~&\sum
    \log
    (
    1
    + {\mathbf{r}_p}
    )
\end{align*}

\noindent
where $\mathbf{r}_p(\mathbf{w}_0 | R^\mathbb{P}, \mathbf{w}_{1}, c,v)=R^\mathbb{P}\mathbf{w}_0^T
- c~v~ \left\|\mathbf{w}_0-\mathbf{w}_1\right\|_1, \mathbf{r}_p \in \mathbb{R}^{N\times 1}
$ is the column vector of possible portfolio returns $r_p$, $c$ denotes the constant transaction, and $v$ is a transaction cost aversion coefficient, which is used to control portfolio turnover.
\\\\
This objective function shares a close relationship with \cref{eq:objfun_ultimate}. Its construction relies on complete distribution information, going beyond first or second-moment statistics. Its primary focus is on maximizing the portfolio's long-term growth rate with no explicit consideration of investor risk aversion. Consequently, this approach can lead to overly aggressive bets. In practice, the objective function can be adapted by truncating at finite moments and adjusting the coefficients associated with $R^\mathbb{P}\odot\mathbf{w}_0$ series to accommodate various risk-return trade-off considerations.
Moreover, by maximizing the geometric mean of portfolio returns instead of the arithmetic mean, the Kelly portfolio mitigates the ``volatility drag" associated with long-term portfolio returns \citep{lo_pursuit_2021}. Implementing the Kelly portfolio may encounter numerical problems when $1+\mathbf{r}_p(\mathbf{w}_0 | R^\mathbb{P}, \mathbf{w}_{1}, c, v) \leq 0$. This issue can be circumvented by confining the search for optimal $\mathbf{w}_0$ in the subspace where $1+\mathbf{r}_p(\mathbf{w}_0 | R^\mathbb{P}, \mathbf{w}_{1}, c)>0$.

\subsubsection{Finite-moments portfolio}
\noindent
The Markowitz-Roy portfolio is formulated using the following general objective function:

\begin{align*}
    \max~&
    \lambda_1 \mathbb{E}[r_p] - \lambda_2 \mathbb{V}[r_p]
    \\=\max~&
    \lambda_1 ({\mathbf{1}} R^{\mathbb{P}} \mathbf{w}_0^T)
    - \lambda_2 
    \big(
    {\mathbf{1}} (R^{\mathbb{P}} \mathbf{w}_0^T)^2
    - ({\mathbf{1}} R^{\mathbb{P}} \mathbf{w}_0^T)^2
    \big)
\end{align*}

\noindent 
where $\mathbf{1} \in \mathbb{R}^{1\times N}$ are row vectors. Several finite-moments-based metrics in this form are proposed in the literature. Given the first and second moments in \gls{mvo}, the maximum entropy distribution is the Gaussian distribution, which allows for exact solutions or accelerated numerical solutions in finite-moment cases. 
\\\\
In this study, we explore a range of finite-moments-based proxy objective functions, targeting portfolio variance (\ref{eq:variance}), expectation (\ref{eq:expretn}), downside frequency (\ref{eq:downsidefreq}), downside variance (\ref{eq:downsidevariance}) \citep{roy_safety_1952}, Sharpe ratio (\ref{eq:sharperatio}) \citep{sharpe_mutual_1966}, Sortino ratio (\ref{eq:sortinoratio}) \citep{sortino_performance_1994}, and Bernado-Ledoit ratio (\ref{eq:bernadoledoitratio}) \citep{bernardo_gain_2000}:

\begin{align}
    % minVariance
    \min~& 
    \mathbb{E}\left[\left(r_p-\mathbb{E}[r_p]\right)^2\right]
    \label{eq:variance}
    % maxExpRetn
    \\\min~&
    -\mathbb{E}[r_p]
    \label{eq:expretn}
    % minDownsideFreq
    \\\min~&
    \mathbb{E}[\mathbb{I}(r_p<0)]
    \label{eq:downsidefreq}
    % minDownsideVariance
    \\\min~&
    \mathbb{E}[r_p^2|r_p<0]
    \label{eq:downsidevariance}
    % maxSharpeRatio
    \\\min~&
    -\log(100+\mathbb{E}[r_p]/\sigma_1)
    \label{eq:sharperatio}
    % maxSortinoRatio
    \\\min~&
    -\log(100+\mathbb{E}[r_p]/\sigma_2)
    \label{eq:sortinoratio}
    % maxBernadoLedoitRatio
    \\\min~&
    \log\left(\mathbb{E}[|r_p|] + \mathbb{E}[|r_p|\big|r_p<0]\right)
    -\log(\mathbb{E}[|r_p|])
    \label{eq:bernadoledoitratio}
\end{align}

\noindent
where ${\mathbf{r}_p} = R^\mathbb{P}\mathbf{w}_0^T - c~v~\left\|\mathbf{w}_0 - \mathbf{w}_1\right\|_1$, $\sigma_1^2=\mathbb{E}\left[(r_p-\mathbb{E}[r_p])^2\right]$ and $\sigma_2^2=\mathbb{E}\left[r_p^2|r_p<0\right]$.

\subsubsection{Quantile portfolio}
\noindent
We also investigate portfolio proxy objective functions targeting \gls{var} \citep{longerstaey_riskmetricstmtechnical_1996} and \gls{es} \citep{rockafellar_optimization_2000} for constructing quantile portfolios:

\begin{align*}
    % minVaR
    \max~&
    \int_0^1F^{-1}_p(u)\delta(u-\alpha)
    ~du
    \\
    % minES
    \max~&
    \int_0^1F^{-1}_p(u)\frac{\mathbb{I}(u\in(0,\alpha))}{\alpha}
    ~du
\end{align*}

\noindent
where $F^{-1}_p(\cdot)$ is the \gls{ppf} of the portfolio return $r_p$, $\delta(\cdot)$ is Dirac's delta function, and $\alpha\in\{0.05, 0.1, 0.5\}$.
\\\\
They are spectral risk measures \citep{acerbi_spectral_2002} that can address the tail part of the $r_p$ distribution. Tail profit/loss events, often challenging to quantify, exert more influence on investment decisions than average profit and loss. As widely accepted tail risk measures, \gls{es} is a coherent risk measure but \gls{var} is not.

\subsubsection{Parity portfolio}
\noindent
Parity portfolios are constructed using heuristic proxy objective functions that target long weight parity (``Talmud") (\ref{eq:eqlong}), short weight parity (\ref{eq:eqshort}), and variance parity (\ref{eq:varparity}) \citep{qian_risk_2011}:

\begin{align}
    % EqualWeightLong
    \max~&
    \cos(\mathbf{w}_0,\mathbf{1})
    \nonumber
    \\=\max~&
    \frac{\mathbf{w}_0}{\|\mathbf{w}_0\|_2}\cdot\mathbf{1}
    \label{eq:eqlong}
    \\
    % EqualWeightShort
    \max~&
    \cos(\mathbf{w}_0,-\mathbf{1})
    \nonumber
    \\=\max~&
    -\frac{\mathbf{w}_0}{\|\mathbf{w}_0\|_2}\cdot\mathbf{1}
    \label{eq:eqshort}
    \\
    % VarParity
    \max~&
    \cos(\mathbf{w}_0\odot\mathbf{\sigma}^2,\mathbf{1})
    \nonumber
    \\=\max~&
    \frac{\mathbf{w}_0\odot\mathbf{\sigma}^2}{\|\mathbf{w}_0\odot\mathbf{\sigma}^2\|_2}\cdot\mathbf{1}
    \label{eq:varparity}
\end{align}

\noindent
where ${\mathbf{\sigma}} \in \mathbb{R}^{1\times D}$ is a row vector filled with the standard deviation of assets returns $\sigma_d=\sqrt{\mathbb{V}({r_d}{w_{0,d}}-c~v~\|w_{0,d}-w_{1,d}\|_1)}$, and $r_d$ denotes the return of asset $d$. Both ${\mathbf{\sigma}\in \mathbb{R}^{1\times D}}$ and ${\mathbf{1} \in \mathbb{R}^{1\times D}}$ have the same shape as ${\mathbf{w}_0 \in \mathbb{R}^{1\times D}}$.

\subsection{Forecasting from generative models}
\noindent
Forecasting is explicitly or implicitly embedded in any portfolio construction scheme. Accurate multivariate asset return dependence modeling is important for diversification in investment portfolios. The generative model approach uses a rolling window of observed historical returns to learn the coherent structure of asset performances over time, and to generate future asset returns. Multivariate elliptical distributions, including multivariate Gaussian and Student's t, are widely accepted in academia and industry for their efficiency and well-understood properties.
Notably, the \gls{dcc}-\gls{garch} model addresses both individual asset volatility clustering and time-varying cross-asset correlations \citep{orskaug_dcc-garch_2009}.
On another front, copula models are designed to disentangle marginal univariate distributions from multivariate dependence patterns \citep{joe_dependence_2014}.
Vine copula models, in particular, have emerged as a versatile tool for constructing multivariate copulae using bivariate building blocks and conditionalizing, allowing for flexible modeling of multivariate asymmetric tail dependencies \citep{czado_vine_2022}.
\\\\
The univariate \gls{garch} $(1,1)$ model is defined as:

\begin{align*}
    r_t=& \mu_t + a_t
    \\
    a_t=& h_t^{1/2} z_t
    \\
    h_t=& \alpha_0 + \alpha_1 a_{t-1}^2 + \beta_1 h_{t-1}
\end{align*}

\noindent
where $r_t$ is the return, $a_t$ is the mean-corrected return, $z_t$ is the standardized error \gls{iid} in $N(0,1)$; $\alpha_0,\alpha_1,\beta_1$ are parameters. The conditional variance $h_t$ is determined based on historical returns. 
%%In some literature, $z_t$ is further fitted using dependence models. 
Forecasts of $r_t$ can be generated from samples of $z_t$ \citep{bollerslev_generalized_1986}.
\\\\
The multivariate \gls{dcc}$(1,1)$-\gls{garch}$(1,1)$ model takes the following form \citep{engle_dynamic_2002}:

\begin{align*}
    \mathbf{r}_t
    &=\mathbf{\mu}_t+\mathbf{a}_t
    \\
    \mathbf{a}_t
    &=\mathbf{z}_t\mathbf{H}_t^{1/2}
    \\
    \mathbf{H}_t
    &=\mathbf{G}_t\mathbf{R}_t\mathbf{G}_t
    \\
    \mathbf{G}_t
    &=\mathrm{diag}{[\sqrt{h_{1,t}}, \dots, \sqrt{h_{d,t}}, \dots, \sqrt{h_{D,t}}]}
    \\
    h_{i,t}
    &=\alpha_{i,0}+\alpha_{i,1}a_{i,t-1}^2+\beta_{i,1}h_{i,t-1}
    \\
    \mathbf{R}_t
    &=\mathbf{Q}_t^{*~-1}\mathbf{Q}_t\mathbf{Q}_t^{*~-1}
    \\
    \mathbf{Q}_t^*
    &=\mathrm{diag}{[\sqrt{q_{11,t}}, \dots, \sqrt{q_{dd,t}}, \dots, \sqrt{q_{DD,t}}]}
    \\
    \mathbf{Q}_t
    &=(1-a-b)\overline{\mathbf{Q}}
    +a \mathbf{\epsilon}_{t-1}^T \mathbf{\epsilon}_{t-1}
    +b {\mathbf{Q}}_{t-1}, 
    ~a \ge 0, b \ge 0, a+b<1
    \\
    \mathbf{\epsilon}_{t}
    &=\mathbf{a}_t\mathbf{G}_t^{-1}
    \sim N(\mathbf{0}, \mathbf{R}_t) 
    \\
    \overline{\mathbf{Q}}
    &=\mathrm{Cov}[\mathbf{\epsilon}_{t}^T \mathbf{\epsilon}_{t}]
    =\mathbb{E}[\mathbf{\epsilon}_{t}^T \mathbf{\epsilon}_{t}],
\end{align*}

\noindent
where ${\mathbf{r}_t} \in \mathbb{R}^{1\times D}$ is a row vector for returns of $D$ assets, ${\mathbf{a}_t} \in \mathbb{R}^{1\times D}$ is a row vector of mean-corrected returns, ${\mathbf{H}_t} \in \mathbb{R}^{D\times D}$ is the \gls{pd} conditional covariance of $\mathbf{a}_t$, ${\mathbf{z}_t} \in \mathbb{R}^{1\times D}$ is a row vector of \gls{iid} errors distributed in multivariate Gaussian or Student's t. ${\mathbf{G}_t} \in \mathbb{R}^{D\times D}$ as a \gls{pd} diagonal matrix has standard deviations from univariate \gls{garch}$(1,1)$ models. ${\mathbf{R}_t} \in \mathbb{R}^{D\times D}$ is the \gls{pd} conditional correlation matrix of the standardized disturbances row vector ${\mathbf{\epsilon}_t}\in \mathbb{R}^{1\times D}$. ${\overline{\mathbf{Q}} \in \mathbb{R}^{D\times D}}$ is the \gls{pd} unconditional covariance matrix of the standardized errors (which can be estimated as time series averages), and ${\mathbf{Q}_t^*}\in \mathbb{R}^{D\times D}$ is a \gls{pd} diagonal matrix with entries from the square root of the diagonal elements of ${\mathbf{Q}_t} \in \mathbb{R}^{D\times D}$. $\mathbf{Q}_t^*$ is designed to rescale elements in $\mathbf{Q}_t$. Forecasts of $\mathbf{r}_t$ can be generated from samples of $\mathbf{z}_t$.
%%{\color{blue}Every \gls{pd} would be a numerical problem.}
\\\\
Dependency modeling using copulas is based on Sklar's theorem, 

\begin{align*}
    F(x_1, \dots, x_D) 
    &=
    C(F_1(x_1), \dots, F_D(x_D))
    \\
    F(F_1^{-1}(u_1), \dots, F_D^{-1}(u_D)) 
    &= 
    C(u_1, \dots, u_D)
\end{align*}

\noindent
where $F$ is the multivariate \gls{cdf} of $X_1,\dots,X_D$, $C$ is the associated copula, $F_d, d\in\{1,\dots,D\}$ are marginal \gls{cdf}s, and $F_d^{-1}$ are marginal \gls{ppf} or the inverse of marginal \gls{cdf} \citep{joe_dependence_2014}.
\\\\
The multivariate Gaussian copula has the form,

\begin{align*}
    C(\mathbf{u};\mathbf{R})
    =
    \Phi_D(\Phi^{-1}(u_1),\dots,\Phi^{-1}(u_D);\mathbf{R})
\end{align*}

\noindent
where $\Phi_D$ and $\Phi$ are the multivariate and univariate Gaussian \gls{cdf}, and $\mathbf{R} \in \mathbb{R}^{D\times D}$ is the \gls{pd} correlation matrix. Simulations of this copula could be done by taking standard Gaussian \gls{cdf} on samples from multivariate Gaussian distribution whose covariance matrix is $\mathbf{R}$. Forecasts of asset returns can then be generated by taking marginal \gls{ppf} \citep{paolella_cobra_2018}.
\\\\
The multivariate Student's t copula has the form,

\begin{align*}
    C(\mathbf{u};\mathbf{R},\nu)
    =
    T_{D,\nu}(T_{1,\nu}^{-1}(u_1),\dots,T_{1,\nu}^{-1}(u_D);\mathbf{R})
\end{align*}

\noindent
where $T_{D,\nu}$ and $T_{1,\nu}$ are the multivariate and univariate Student's t \gls{cdf}s with the same degree of freedom $\nu$, and $\mathbf{R} \in \mathbb{R}^{D\times D}$ is the \gls{pd} correlation matrix. A similar procedure can be used to generate asset return forecasts from this copula. 
\\\\
The regular vine copulae are constructed recursively. For every pair of bivariate copulae data, we fit pair copula from elliptical or Archimedean families by \gls{mle} and select the one with minimal \gls{aic} \citep{akaike_information_1998}. The regular vine copula structure is constructed by picking the maximum spanning tree, which maximizes the sum of absolute values of Kendall's $\tau$ among pairwise variables while adhering to proximity conditions \citep{dismann_selecting_2013}.
%%{\color{blue}\\(Structure of D-Vine is fixed once the level 0 tree is known, but C-Vine needs to do MST multiple times. or we just mention this R-Vine construction method as 'the Dissmann algorithm': always use proximity conditions, and do MST whenever necessary.)}
Simulations of this copula could be done by performing inverse Rosenblatt transform \citep{rosenblatt_remarks_1952} on \gls{iid} multivariate uniform samples \citep{czado_analyzing_2019}. Forecasts of asset returns can then be generated by taking marginal \gls{ppf}.
\\\\
In this study, we cover a diverse set of generative models for asset returns forecasting. The asset returns in historical windows, available at every rebalancing step, are used as inputs to fit generative models.
We fit multivariate Gaussian and Student's t distributions, multivariate Gaussian and Student's t copulae, and regular-vine copulae using combinations of elliptical family copulae and Archimedean family copulae. We build \gls{dcc}$(1,1)$-\gls{garch}$(1,1)$ models using Gaussian and Student's t-distributed errors. To address temporal serial correlation before modeling dependency across assets, we also include models where residuals from univariate \gls{garch}$(1,1)$ models are fed to Gaussian, Student's t, and Vine copulae. Both parametric and non-parametric (empirical) univariate marginal \gls{cdf} and \gls{ppf} are used to process returns into copula data. We consider a range of candidate parametric marginal \gls{cdf}s including Gaussian, Student's t, non-central Student's t, Johnson's SU, Tukey-lambda, Laplace, and asymmetric Laplace distributions. Fitted marginals are selected to minimize \gls{aic}. All parametric models are fitted by \gls{mle} using historical observed data.

\subsection{Backtest}\label{sec:fix_backtest}
\begin{figure}
    \centering
    \includegraphics[width=.8\textwidth]{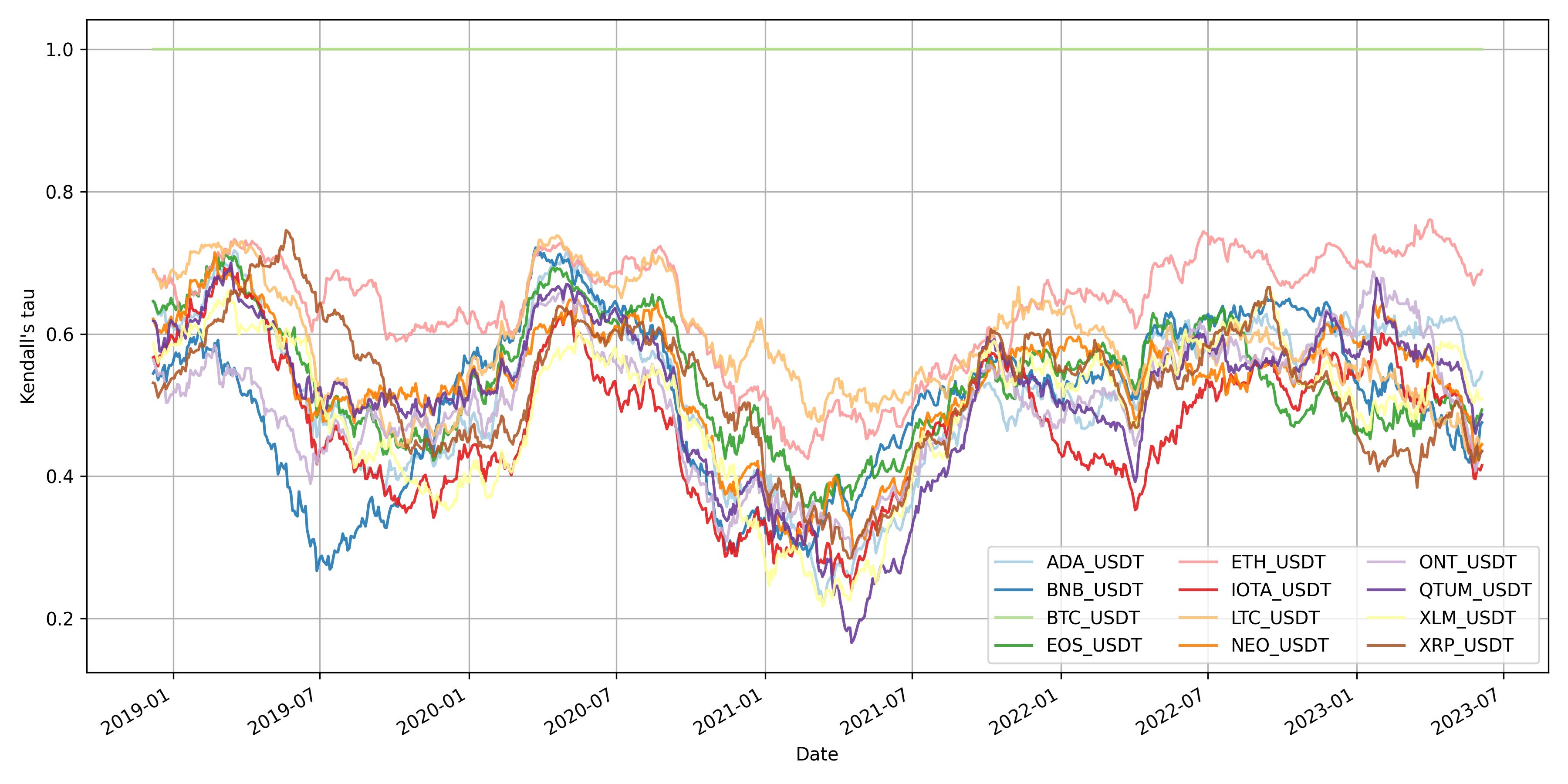}
    \caption{Kendall's $\tau$ between BTC and studied cryptocurrencies, using a $182$-day rolling window of $2$-day simple returns as priced in USDT on Binance.}
    \label{fig:fix_tau}
\end{figure}

\noindent
To backtest generative model-based portfolios, we retrieved price time series from Binance, spanning from 2018-06-10 to 2023-06-04, covering $12$ currencies priced in USDT including ADA, BNB, BTC, EOS, ETH, IOTA, LTC, NEO, ONT, QTUM, XLM, and XRP. As depicted in \cref{fig:fix_tau}, asset returns have notable high correlations. Rebalancing occurs every two days, targeting diverse combinations of generative dependence model forecasts and proxy objective functions. To fit the generative model, we employ a rolling window length of $91$ steps or $182$ days. The transaction cost $c$ is set as $50$ bps, together with the number of assets $D=12$ and the constraint boundary multiplier $m=5$. To ensure robustness and assess performance consistency, we conduct multiple independent simulations using different random seeds. In total, we have $6525$ simulation paths, each comprising $820$ steps.
\\\\
For performance attribution, we fit the \gls{lasso} model (\ref{eq:lasso}) and analyze portfolio performance at each step \citep{efron_least_2004,pedregosa_scikit-learn_2011}. We use the portfolio simple return $r_p$ (\ref{eq:rp}) and the logit of cosine similarity between $\mathbf{w}_{0,t}$ and $\mathbf{r}_t$ (\ref{eq:logit_cosine}) as the performance measure $y$. 
In \cref{eq:logit_cosine}, the cosine similarity is used to quantify decision quality, and the logit transform extends its value from $(0,1)$ to the entire real line. The whole logit-cosine performance measure is designed to be a consistent ex-post metric, enabling comparisons of simulation outcomes at different steps from different generative model / objective function setups. The \gls{lasso} model, known for its capacity to produce sparse yet robust coefficients, is employed primarily for its interpretable coefficient ranking.

\begin{align}
    \min~&
    \frac{1}{2~n_{\text{samples}}} \|\mathbf{y}-\mathbf{X}\hat{\beta}\|_2^2 + \lambda~\|\hat{\beta}\|_1
    \label{eq:lasso}
    \\
    y=&~r_p
    \label{eq:rp}
    \\
    y=&~
    \text{logit}\left(\frac{1+\cos(\mathbf{w}_{0,t}, \mathbf{r}_{t})}{2}\right)
    \label{eq:logit_cosine}
    % \\
    % y=&
    % \text{logit}\left(\frac{\|\mathbf{w}_{0,t}-\mathbf{w}_{1,t-1}\|_1}{2}\right)
    % \label{eq:logit_turnover}
    % \\
    % y=&
    % \text{logit}\left(\frac{1}{D~\|\mathbf{w}_{1,t}\|_2^2}\right)
    % \label{eq:logit_numeffw1}
\end{align}

\begin{figure}
    \centering
     \begin{subfigure}[b]{0.48\textwidth}
         \centering
         \includegraphics[width=\textwidth]{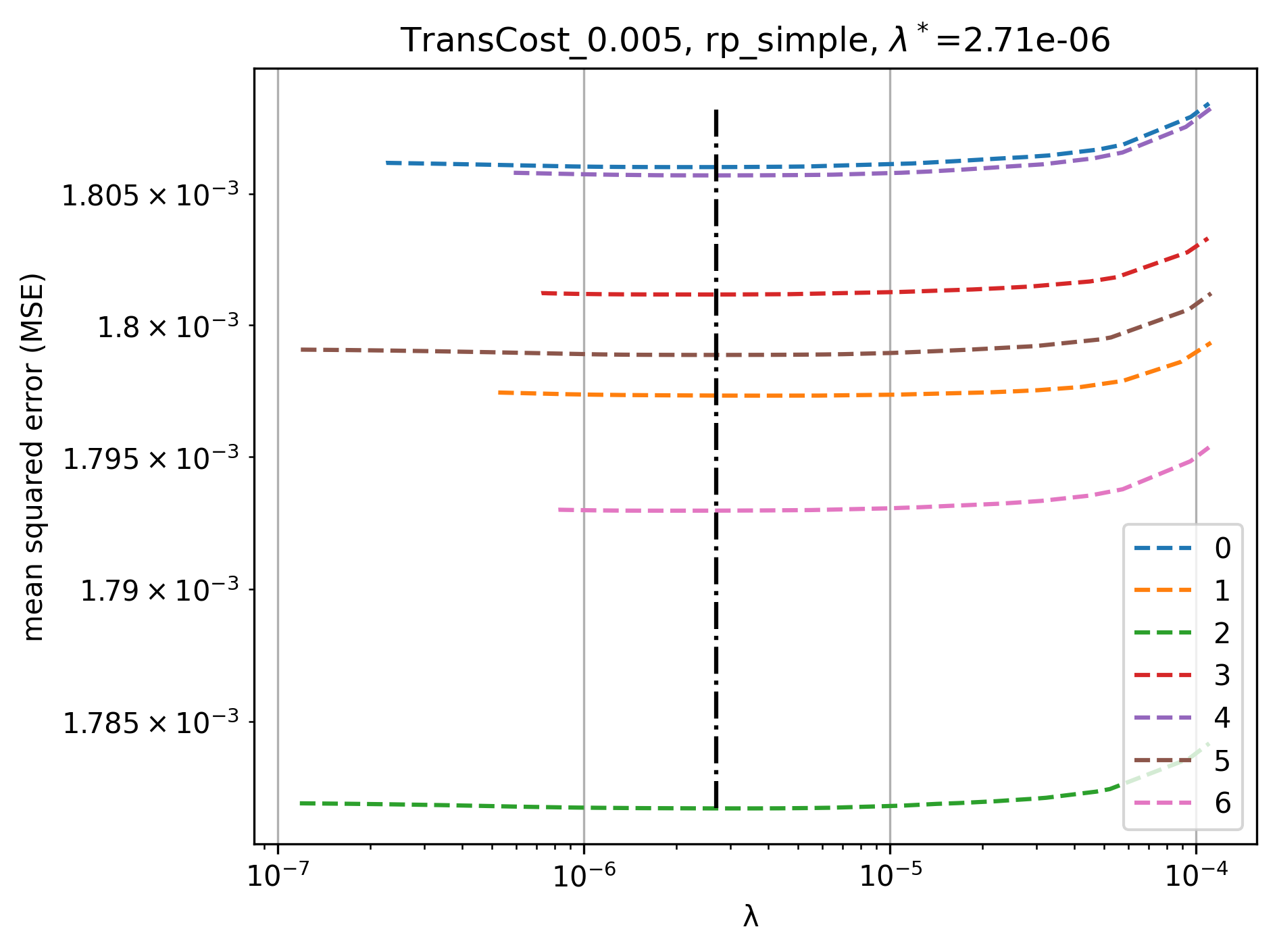}
         \caption{Mean squared error in each fold.}
         \label{fig:fix_lasso_path_cv_rp}
     \end{subfigure}
     \hfill
     \begin{subfigure}[b]{0.48\textwidth}
         \centering
         \includegraphics[width=\textwidth]{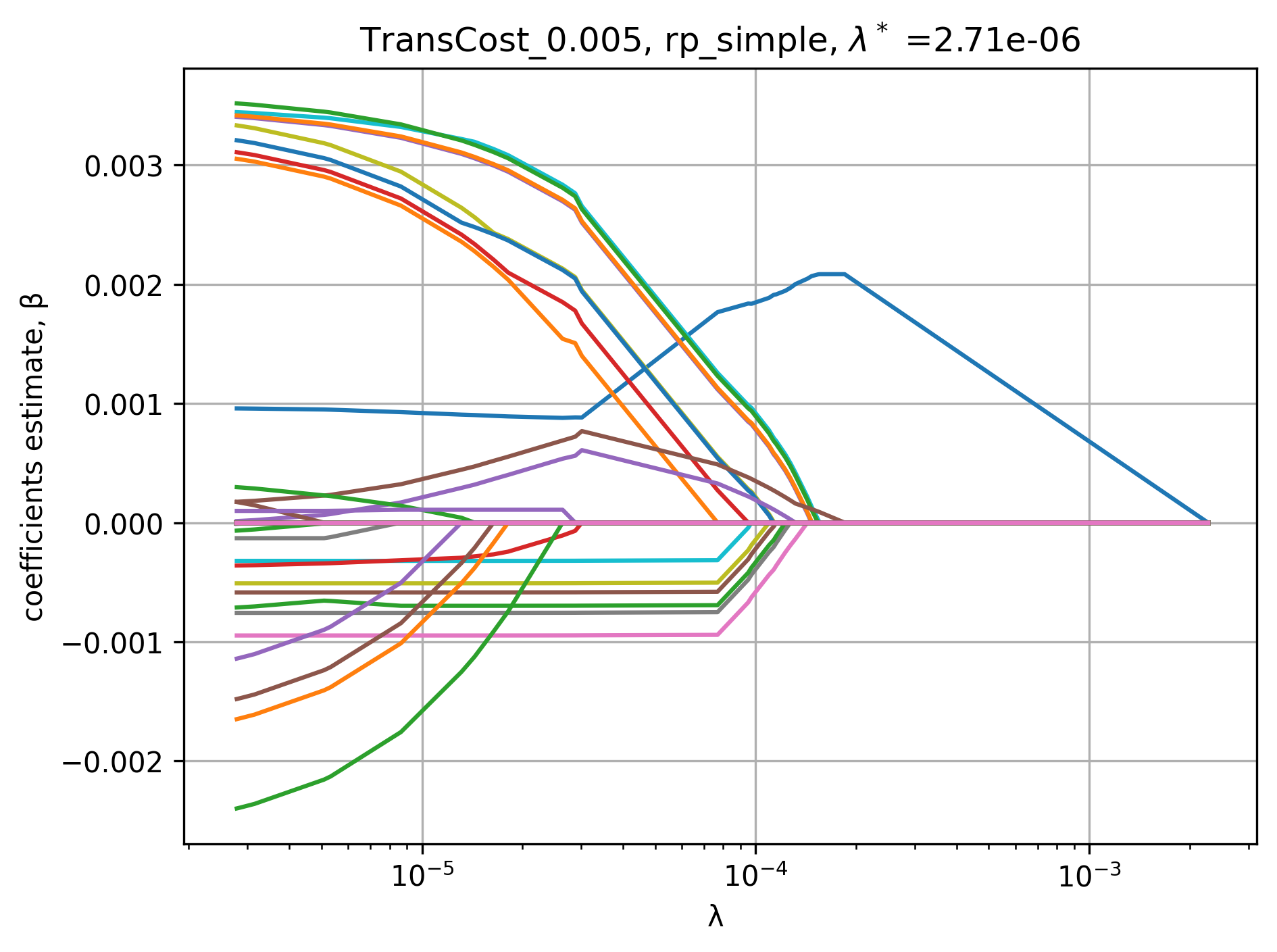}
         \caption{Coefficients in the final \gls{lasso}.}
         \label{fig:fix_lasso_path_coef_rp}
     \end{subfigure}
     \caption{Cross-validated \gls{lasso} paths for portfolio return $r_p$ attribution. The optimal regularization strength hyperparameter $\lambda^*$ is the average of $\lambda$ where mean squared errors are minimized in each fold.}
\end{figure}

\noindent
The design matrix ${\mathbf{X}}\in \mathbb{R}^{5350500\times 597}$ is filled with binary values ($0$ or $1$) encoding the presence of foundation elements for portfolio construction: the intercept, various generative models, objective functions, the transaction cost aversion coefficient $v$, and specific combinations of them (as interaction terms indicating the simultaneous presence of the underlying elements). While $\mathbf{X}$ possesses $597$ columns, each row contains only seven instances of the value $1$. The performance measures, represented as the column vector ${\mathbf{y}}\in \mathbb{R}^{5350500\times 1}$ encompass data from all steps in all simulation paths. 
As the scalar regularization strength hyper-parameter $\lambda$ increases, sparser ${\hat{\beta}}\in \mathbb{R}^{597\times 1}$ estimates are obtained. The regularization process results in the shrinkage of coefficients for unimportant covariates, effectively reducing them to $0$. Any remaining influence from these unimportant covariates is typically aggregated within the intercept term. The optimal $\lambda^*$ is selected based on $7$-fold \gls{cv} that minimizes \gls{mse}. For illustration, we visualize the \gls{mse} paths in each fold during \gls{cv} for $r_p$ in \cref{fig:fix_lasso_path_cv_rp} and coefficient paths during the final fit using all observations for $r_p$ in \cref{fig:fix_lasso_path_coef_rp}.

\begin{figure}
    \centering
    \includegraphics[width=.8\textwidth]{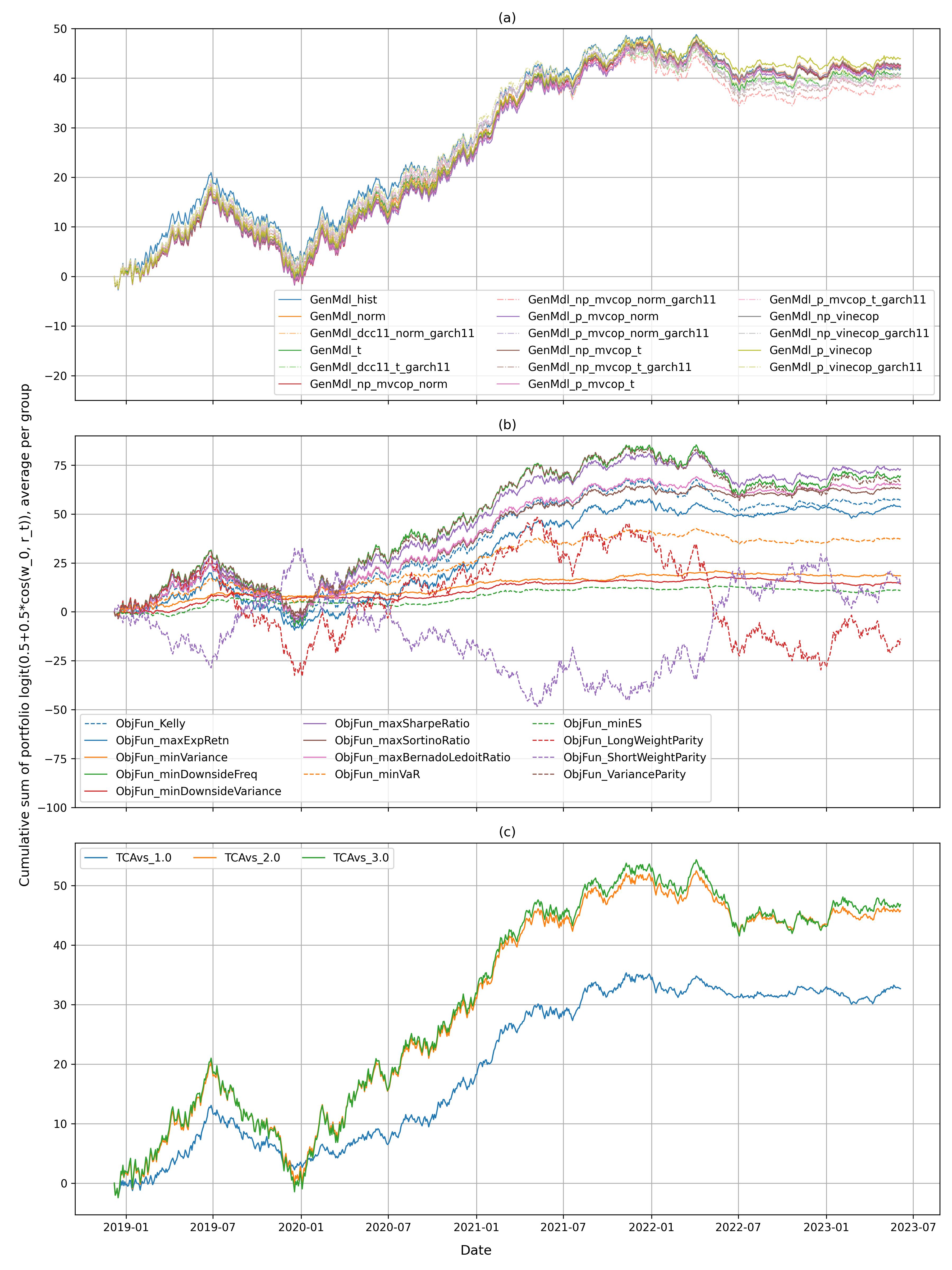}
    \caption{Cumulative sum of logit-cosine performance measure, averaged for portfolios grouped by (a) asset returns generative models, (b) proxy objective functions and (c) transaction cost aversion coefficients.}
    \label{fig:fix_cumsum_logitcos}
\end{figure}

\noindent\\
The cumulative sum of logit-cosine performance measure is visualized in \cref{fig:fix_cumsum_logitcos}. The symmetry of the logit-cosine performance measure is shown in \cref{fig:fix_cumsum_logitcos}b where lines representing the long-weight parity portfolio and the short-weight parity portfolio are mirrored around the horizontal axis.
We present the non-zero coefficients obtained at the optimal $\lambda^*$ in \gls{lasso} for the $r_p$ in \cref{tab:fix_lasso_coef_rp} and for the logit-cosine performance measure in \cref{tab:fix_lasso_coef_logit_cosine}. 

\begin{table}[!ht]
\centering\scriptsize
\caption{
Non-zero coefficients in \gls{lasso} for portfolio simple return, corresponding to combinations of proxy objective functions, generative models, and transaction cost aversion coefficients.
}
\label{tab:fix_lasso_coef_rp}
\begin{tabular}{lr}
\toprule
coefficient & value \\
\midrule
ObjFun maxSharpeRatio & 0.003516 \\
ObjFun VarianceParity & 0.003443 \\
ObjFun minVaR 0.5 & 0.003415 \\
ObjFun minDownsideFreq & 0.003403 \\
ObjFun Kelly & 0.003332 \\
ObjFun maxBernadoLedoitRatio & 0.003207 \\
ObjFun maxSortinoRatio & 0.003107 \\
ObjFun maxExpRetn & 0.003051 \\
intercept & 0.000958 \\
ObjFun VarianceParity : TCAvs 1.0 & 0.000297 \\
TCAvs 3.0 & 0.000175 \\
ObjFun minVariance : TCAvs 1.0 & 0.000173 \\
ObjFun maxExpRetn : TCAvs 3.0 & 0.000098 \\
TCAvs 2.0 & 0.000013 \\
GenMdl p vinecop archimedean & 0.000002 \\
GenMdl np vinecop garch11 elliptical & -0.000010 \\
GenMdl dcc11 t garch11 & -0.000067 \\
ObjFun minVariance : TCAvs 3.0 & -0.000131 \\
ObjFun minVaR 0.05 & -0.000321 \\
TCAvs 1.0 & -0.000360 \\
ObjFun minES 0.5 & -0.000509 \\
ObjFun minDownsideVariance & -0.000585 \\
ObjFun minVariance & -0.000712 \\
ObjFun minES 0.1 & -0.000757 \\
ObjFun minES 0.05 & -0.000947 \\
ObjFun maxBernadoLedoitRatio : TCAvs 1.0 & -0.001141 \\
ObjFun Kelly : TCAvs 1.0 & -0.001479 \\
ObjFun maxSortinoRatio : TCAvs 1.0 & -0.001648 \\
ObjFun maxExpRetn : TCAvs 1.0 & -0.002397 \\
\bottomrule
\end{tabular}
\end{table}

\begin{figure}
    \centering
    \includegraphics[width=.8\textwidth]{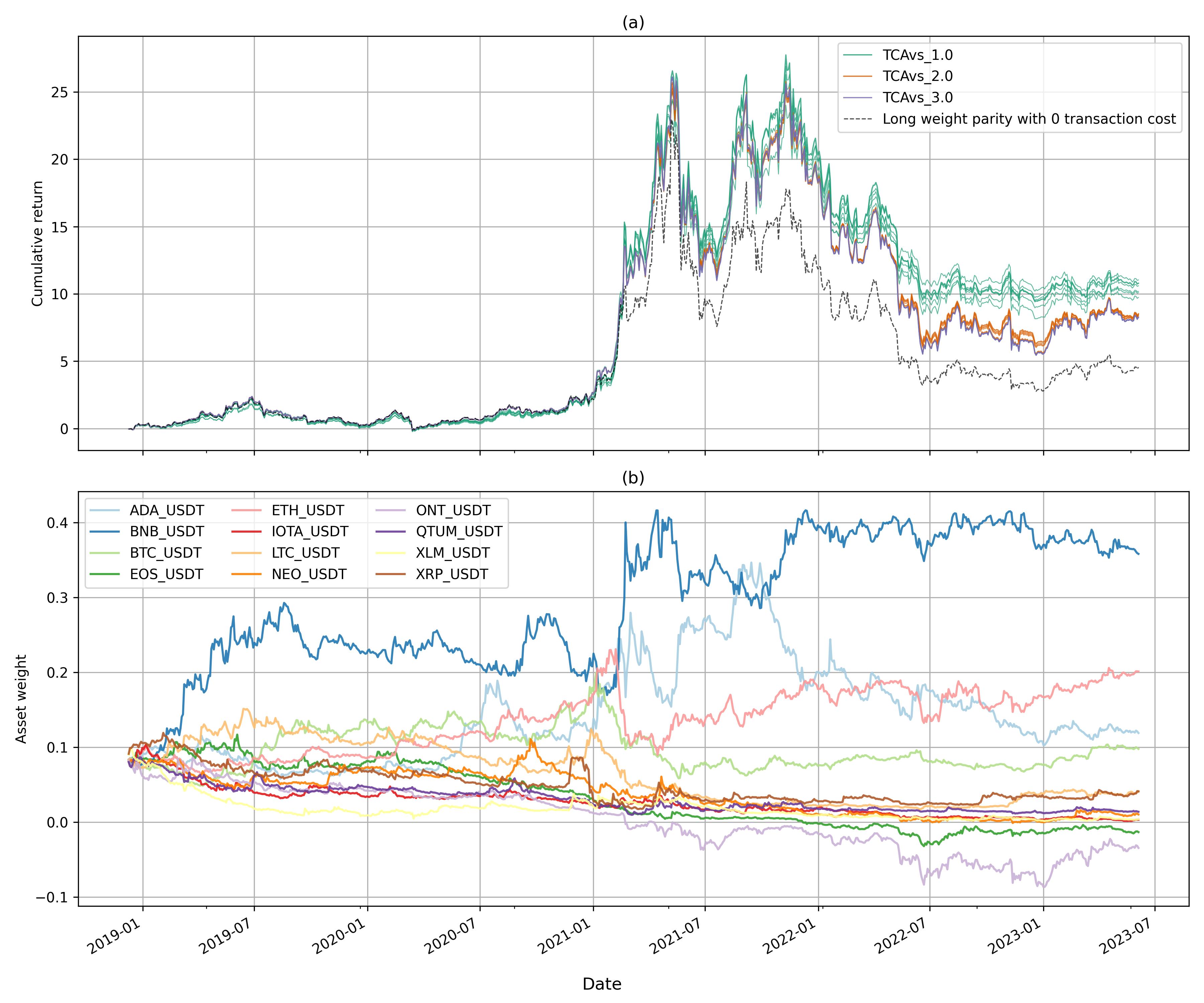}
    \caption{Visualization of portfolios using vine copula with parametric univariate marginal distributions and Archimedean bivariate copulae as asset returns forecasting generative models along with Sharpe ratio maximization as proxy objective functions, in (a) cumulative simple returns and (b) averaged asset weights.}
    \label{fig:fix_best}
\end{figure}

\begin{table}[!ht]
\centering\scriptsize
\caption{
Non-zero coefficients in \gls{lasso} for the logit-cosine performance measure, corresponding to combinations of proxy objective functions, generative models, and transaction cost aversion coefficients.
}
\label{tab:fix_lasso_coef_logit_cosine}
\begin{tabular}{lr}
\toprule
coefficient & value \\
\midrule
ObjFun maxSharpeRatio & 0.058424 \\
ObjFun maxBernadoLedoitRatio & 0.054276 \\
ObjFun minDownsideFreq & 0.054091 \\
ObjFun minVaR 0.5 & 0.051835 \\
ObjFun maxExpRetn : TCAvs 2.0 & 0.050512 \\
ObjFun Kelly & 0.049198 \\
ObjFun maxExpRetn : TCAvs 3.0 & 0.047490 \\
ObjFun VarianceParity & 0.047314 \\
ObjFun maxSortinoRatio : TCAvs 2.0 & 0.034248 \\
intercept & 0.031102 \\
ObjFun maxSortinoRatio : TCAvs 3.0 & 0.028144 \\
ObjFun maxSortinoRatio & 0.025601 \\
ObjFun VarianceParity : TCAvs 1.0 & 0.013741 \\
ObjFun minVaR 0.5 : TCAvs 1.0 & 0.007737 \\
TCAvs 3.0 & 0.002453 \\
GenMdl p vinecop archimedean & 0.002403 \\
ObjFun maxBernadoLedoitRatio : TCAvs 2.0 & 0.002340 \\
ObjFun maxExpRetn & 0.002289 \\
ObjFun Kelly : TCAvs 2.0 & 0.001262 \\
GenMdl p vinecop allfam & 0.000581 \\
ObjFun minDownsideFreq : TCAvs 1.0 & 0.000114 \\
GenMdl dcc11 t garch11 & -0.000200 \\
GenMdl p mvcop t garch11 & -0.000259 \\
GenMdl np vinecop garch11 allfam & -0.000993 \\
ObjFun minDownsideVariance : TCAvs 1.0 & -0.001419 \\
GenMdl np mvcop norm garch11 & -0.002232 \\
ObjFun minVariance & -0.004122 \\
ObjFun minVaR 0.05 & -0.004588 \\
ObjFun minVariance : TCAvs 3.0 & -0.004930 \\
ObjFun minES 0.5 & -0.006759 \\
TCAvs 1.0 & -0.007696 \\
ObjFun minDownsideVariance & -0.010017 \\
ObjFun minES 0.1 & -0.016901 \\
ObjFun maxBernadoLedoitRatio : TCAvs 1.0 & -0.018773 \\
ObjFun minES 0.05 & -0.020972 \\
ObjFun Kelly : TCAvs 1.0 & -0.030433 \\
\bottomrule
\end{tabular}
\end{table}

\noindent\\
In general, during the periods we considered, the cryptocurrency market (priced in USDT) exhibited an upward trend as indicated by the positive intercept term in both tables. The choice of the proxy objective function is shown to have a greater impact as evidenced by the larger magnitude of their coefficients in both tables. Among the generative models, only the vine copula yields positive coefficients in both tables, emphasizing its superiority in multivariate dependence modeling. It's worth noting that interaction terms corresponding to the presence of specific combinations of the generative model and objective function have shrunken to $0$. This suggests that it is reasonable to divide the portfolio construction task into asset return forecasting and optimizations. The Sharpe ratio portfolios, as classical finite-moments portfolios, outperformed Kelly's portfolios, quantile portfolios, and parity portfolios. 
\\\\
To illustrate the best combination suggested by the \gls{lasso} results, we chart the cumulative returns of portfolios using parametric marginal distribution and vine copulas based on Archimedean bivariate copula as the generative model paired with Sharpe ratio as the objective function in \cref{fig:fix_best}a. Their averaged asset weights are depicted in \cref{fig:fix_best}b. Notably, all simulated paths generated using this combination outperformed the benchmark, which adheres to the long-weight parity objective function but with no transaction costs included in the portfolio return. A moderate transaction cost aversion coefficient $v$ around $2$ to $3$ is favorable, but too high a hurdle for trading could limit terminal profits. These portfolios often contain a strong long position in BNB, ETH, ADA, and BTC, and even reach the upper weight limit $m/D$ for BNB occasionally.

\section{Constructing eclectic portfolios: three essential components}
\label{sec:eclectic}
\noindent
In this session, we expand our exploration beyond the confines of a single combination of proxy objective functions and generative model forecasts. We delve into the realm of portfolio blending, embracing diverse portfolio theories and strategies. Our approach, as outlined in \cref{sec:genport}, involves an extension of the \gls{mab} framework. 
\\\\
In the classical \gls{mab} Markov decision process, a decision-maker seeks to maximize rewards obtained from a limited number of lever pulls on slot machines. In other words, the agent interacts with a one-state environment and searches for the best action \citep{sutton_reinforcement_2018,thompson_likelihood_1933}. 
In our context, the agent's action space corresponds to the set of arms representing different portfolio choices, and the environment's rewards associated with each arm are determined by the performance of the corresponding portfolio. 
The agent holds both a value model to evaluate the relative optimality among arms, and also a policy model to decide which arm to pull.
\\\\
Eclectic portfolios are constructed by allocating blending ratios ${\psi} \in \mathbb{R}^{1\times P}$ to portfolios of different investment styles. At each step, the agent updates the fractional blending ratio $\psi_p$ for each portfolio $p$, based on their historical performances up to that point. In other words, we work towards \cref{eq:objfun_ultimate} by blending optimal weights solved from different one-period \cref{eq:objfun} in a long-only manner ($\|\psi\|_1=1,~\psi_p>0$). The temporal coherence exhibited in the market dynamics and portfolio performances allows such reinforcement learning to explore better portfolio blending. 
\\\\
Three components are essential: similarity, optimality, and blending ratio. Similarity, denoted as $s(\cdot)$, defines comparable portfolio performance metrics. Optimality quantifies arm preferences among peers using activation functions $\pi(\cdot)$. Blending ratios $\psi(\cdot)$ determine the allocation strategy. In essence, $s(\cdot)$ and $\pi(\cdot)$ constitute the value model, while $\psi(\cdot)$ encodes the policy model.
\\\\
Practical principles are embedded in this reinforcement learning task. Better-performing arms should have higher $s(\cdot)$ and $\pi(\cdot)$ in value models. Since the market is adaptive \citep{lo_adaptive_2017}, recent performances should carry more weight in policy models.

%%{\color{red}\\workflow:$\mathbf{w}_0,\mathbf{r}\overset{sim~func}{\rightarrow} s \overset{act~func}{\rightarrow} g \overset{1-norm~rescale}{\rightarrow} \pi \overset{par.est.}{\rightarrow} \theta,\gamma \overset{greedy~or~prob.mat}{\rightarrow} \psi$}

\subsection{Similarity}
\noindent
The initial step involves reviewing recent portfolio weights to compute comparable performance statistics across various portfolios. This measure should highlight decision quality, be comparable at different time steps, be positively correlated to portfolio returns, and not be specific to $\mathbf{w}_{1,t-1,p}$, $c$, and $v$. 
\\\\
We calculate the ex-post similarity row vector $\mathbf{s}_{t} \in \mathbb{R}^{1\times P}$ by assessing how closely the realized asset returns $\mathbf{r}_{t}$ align with the solutions $\mathbf{w}_{0,t,p}$ obtained from \cref{eq:objfun}, where $t$ denotes the time step and $P$ represents the number of arms or portfolios.
\\\\
The cosine similarity of portfolio $p\in\{1,2,\dots,P\}$ is the cosine of the angle between $\mathbf{w}_{0,t,p} \in \mathbb{R}^{1\times D}$ and ${\mathbf{r}_t} \in \mathbb{R}^{1\times D}$:

\begin{align*}
    s_{\cos,t,p}=
    &\frac{
    \mathbf{w}_{0,t,p}
    \cdot
    \mathbf{r}_t    
    }
    {
    \|\mathbf{w}_{0,t,p}\|_2
    ~
    \|\mathbf{r}_t\|_2
    }
\end{align*}

\noindent
The Z-score similarity is defined as:
\begin{align*}
    s_{Z,t,p}=
    &2~\Phi(\mathbf{w}_{0,t,p} \cdot \mathbf{r}_t)-1
\end{align*}

\noindent
where $\Phi(\cdot)$ is the \gls{cdf} of standard Gaussian distribution.
\\\\
Other similarities based on the vector norms of $\mathbf{d}_{t,p}=\frac{\mathbf{w}_{0,t,p}}{\|\mathbf{w}_{0,t,p}\|_1}-\frac{\mathbf{r}_t}{\|\mathbf{r}_t\|_1}$ (which is the distance between two points on an $L1$ unit sphere) are also used:

\begin{align*}
    s_{L1,t,p}=&1-\|\mathbf{d}_{t,p}\|_1
    \\
    s_{L2,t,p}=&1-\|\mathbf{d}_{t,p}\|_2
    \\
    s_{L{\infty},t,p}=&1-\|\mathbf{d}_{t,p}\|_\infty
\end{align*}

\noindent
All of the above measures use re-scaled returns to focus on the decision quality of an arm. These measures are bounded inside $[-1,1]$ and are strongly correlated with portfolio returns.

\subsection{Optimality}
\noindent
The second step is to assess the optimality of each portfolio $p$ among peers $\pi_{t,p}$, by applying activation functions to similarities.
\\\\
The naive greedy activation function is the maxout \cref{eq:maxout}, which compares the similarities of different portfolios at the same time step $t$, then assigns a grade of $1$ to those with the highest $s(\cdot)$ and $0$ to the rest. We rescale grades ${\mathbf{g}_t} \in \mathbb{R}^{1\times P}$ of portfolios $p$ by sum to get $\mathbf{\pi}_{t,p}$ if there are ties. 

\begin{align}
    g_{t,p}=&
    \mathbb{I}\{p=\arg\max_q({s}_{t,q})\}
    \label{eq:maxout}
    \\
    \pi_{t,p}=&
    \frac{g_{t,p}}{\|\mathbf{g}_t\|_1}
    \nonumber
\end{align}

\noindent
The optimality scores are organized in a matrix, where the column indexes a portfolio $p$ and the row indexes a previous step $t$. If there are multiple strategies to evaluate, the matrix $\mathbf{\pi}_{t,p}$ filled with the maxout activation function will be wide and sparse. This study also explores non-greedy activation functions including softmax (\ref{eq:softmax}), logistic (\ref{eq:logistic}), tanh (\ref{eq:tanh}), 
% identity (\ref{eq:identity}),
leaky-relu (\ref{eq:relu}), logit (\ref{eq:logit}), and probit (\ref{eq:probit}):

\begin{align}
    g_{t,p}=&
    \exp(7{s}_{t,p})
    \label{eq:softmax}
    \\
    g_{t,p}=&
    \frac{1}{1+\exp(-7{s}_{t,p})}
    \label{eq:logistic}
    \\
    g_{t,p}=&
    1+\tanh(7{s}_{t,p})
    \label{eq:tanh}
    % \\
    % g_{t,p}=&
    % 1+{s}_{t,p}
    % \label{eq:identity}
    \\
    g_{t,p}=&
    1/7
    +
    7s_{t,p}\mathbb{I}\{s_{t,p}\geq0\}
    -
    7s_{t,p}\mathbb{I}\{s_{t,p}<0\}
    \label{eq:relu}
    \\
    g_{t,p}=&
    \max\left(0, \text{logit}\left(\frac{1+{s}_{t,p}}{2}\right)\right)
    \label{eq:logit}
    \\
    g_{t,p}=&
    \max\left(0,\Phi^{-1}\left(\frac{1+{s}_{t,p}}{2}\right)\right)
    \label{eq:probit}
    \\\nonumber
    \\
    \mathbf{\pi}_{t,p}=&
    \frac{g_{t,p}}{\|\mathbf{g}_t\|_1}
    \nonumber
\end{align}

\noindent
where $\Phi^{-1}(\cdot)$ is the \gls{ppf} of standard Gaussian distribution. All of the above $\pi_{t,p}$ are bounded inside $[0,1]$ and give higher values to those with higher similarities.

\subsection{Blending or switching: to randomize or to maximize}
\noindent
In the third step, portfolios with higher recent optimality among peers $\pi_p$ are assigned higher action preferences $\psi_p$. As the agent's policy model, we fit distributions for $\pi_{p}$, to estimate their parameters $\theta_{p}$ and calculate the blending ratio $\psi_p$.

\subsubsection{Blending}
\noindent
Portfolio blending is achieved through a probabilistic approach known as probability-matching or Herrnstein's law \citep{lo_maximize_2021}, where we estimate the parameters ${\mathbf{\theta}} \in \mathbb{R}^{1\times P}$ by assuming multivariate distributions for $\mathbf{\pi}$ and then allocate fractional blending weights $\psi_{p}\in(0,1)$ proportional to $\mathbb{E}[\mathbf{\pi}_{p}|\mathbf{\theta}]$:

\begin{align*}
    \psi_{p}=&
    \frac{\mathbb{E}[\mathbf{\pi}_{p}|\mathbf{\theta}]}
    {\sum_p\mathbb{E}[\mathbf{\pi}_{p}|\mathbf{\theta}]}
\end{align*}

\noindent
For $\pi_p$ generated from the maxout activation function, we fit the categorical distribution, $\textrm{Cat}(\theta_1,\dots,\theta_p,\dots,\theta_P)$ using \gls{wmle}:
% https://math.stackexchange.com/questions/2725539/maximum-likelihood-estimator-of-categorical-distribution

\begin{align*}
    \hat{\theta}=~
    &\arg\max
    \sum_p\sum_t
    \gamma^{t}\mathcal{L}(\theta_p|\pi_{t,p})
    \\=~
    &\arg\max
    \sum_p \sum_t \gamma^{t} \pi_{t,p} \log(\theta_{p})
    \\s.t.~ 
    & \| \theta \|_1=1, \theta_p>0
    \\
    \hat{\theta}_p=~
    &\frac{
    \sum_t \gamma^t \pi_{t,p}
    }{
    \sum_p\sum_t \gamma^t \pi_{t,p}
    }
\end{align*}

\noindent
where the sum is over the previous $t$ steps in the rolling window and the portfolios are indexed by $p$. $\gamma\in(0,1)$ is the decay factor that places more weight on recent performances, and $\hat\theta_p$ is the parameter estimate of $\pi_{p}$ after reviewing a rolling window.
\\\\
For $\pi_p$ generated from other activation functions, we fit the Dirichlet distribution, $\textrm{Dir}(\theta_1,\dots,\theta_p,\dots,\theta_P)$ using \gls{wmle}: 
% https://stats.stackexchange.com/questions/130329/maximum-likelihood-estimation-of-dirichlet-mean

\begin{align*}
    \hat{\theta}=~
    &\arg\max
    \sum_t
    \gamma^{t}\mathcal{L}(\theta_p|\pi_{t,p})
    \\=~
    &\arg\max
    \sum_t \gamma^t\log\left[
    \frac{\Gamma(\sum_p \theta_p)}{\prod_p \Gamma(\theta_p)}\prod_p{\pi_{t,p}^{\theta_p -1}}
    \right]
    \\s.t.~ 
    & \theta_p>0
\end{align*}

\noindent
where $\Gamma(\cdot)$ is the gamma function.
\\\\
For both the categorical distribution and Dirichlet distribution, we have $\mathbb{E}[\mathbf{\pi}_{p}|\mathbf{\theta}_{p}]=\theta_p$. Thus, the blending ratio vector $\psi$ and the parameter estimates vector $\hat{\theta}$ are aligned,

\begin{equation*}
    \hat{\psi}_p=
    \frac{\hat{\theta}_p}{\sum_p \hat{\theta}_p}
\end{equation*}

\noindent
Note the probability-matching strategy can also be conceptualized as maximizing the projection from the 2-norm re-scaled blending ratio vector $\psi$ onto the parameter estimates vector $\hat{\theta}$.

%%{\color{blue} Intuitively this is to align the decision vector $\psi$ to the rewards vector $\hat{\theta}$. The previous 1-norm constraint is a suitable generalization of the long-only constraint here.}

\begin{align*}
    \hat{\psi}=~
    &\arg\max~
    \cos(\psi, \hat{\theta})
    \\=~
    &\arg\max~
    \frac{\psi}{\| \psi\|_2}\cdot\hat{\theta}
    \\s.t.~ 
    & \| \psi \|_1=1, \psi_p>0
\end{align*}

\subsubsection{Switching}
\noindent
Portfolio switching, as a special case of blending, restricts binary $\psi_p\in\{0,1\}$, indicating that investors hold strong positive beliefs and invest entirely in the arm with the highest $\theta_p$. Switching is often rationalized in high-conviction investment scenarios, reflecting the belief that ``wide diversification is only required when investors do not understand what they are doing". In the case of portfolio switching, $\pi_p$ can be modeled as a univariate random variable for each $p$. 
\\\\
For $\pi_p$ generated from the maxout activation function, we fit a Bernoulli distribution, $\textrm{Bernoulli}(\theta_p)$ for each portfolio $p$ using \gls{wmle}:
% https://stats.stackexchange.com/questions/275380/maximum-likelihood-estimation-for-bernoulli-distribution

\begin{align*}
    \hat{\theta}=~
    &\arg\max
    \sum_p\sum_t
    \gamma^{t}\mathcal{L}(\theta_p|\pi_{t,p})
    \\
    \hat{\theta}_p=~
    &\arg\max
    \sum_t \gamma^t \Big(
    \pi_{t,p}\log\theta_p
    + 
    (1-\pi_{t,p})\log(1-\theta_p)
    \Big)
    \\s.t.~ 
    &\theta_p\in(0,1)
    \\
    \hat{\theta}_p=~
    &\frac{
    \sum_t\gamma^t\pi_{t,p}
    }{
    \sum_t\big(\gamma^t\pi_{t,p}+\gamma^t(1-\pi_{t,p})\big)
    }
    =\frac{
    \sum_t\gamma^t\pi_{t,p}
    }{
    \sum_t\gamma^t
    }
\end{align*}

\noindent
For $\pi_p$ generated from other activation functions, we fit a beta distribution, $\textrm{Beta}(\theta_p,\nu_p)$ for each portfolio $p$ using \gls{wmle}: 
% https://math.stackexchange.com/questions/2649775/mle-maximum-likelihood-estimator-of-beta-distribution

\begin{align*}
    \hat{\theta}=~
    &\arg\max
    \sum_p\sum_t
    \gamma^{t}\mathcal{L}(\theta_p|\pi_{t,p})
    \\
    \hat{\theta}_p=~
    &\arg\max
    \sum_t \gamma^t \Big(
    (\theta_p\nu_p-1)\log(\pi_{t,p})
    +(\nu_p-\theta_p\nu_p-1)\log(1-\pi_{t,p})
    \\&~~~~
    -\log\mathcal{B}(\theta_p\nu_p,~\nu_p-\theta_p\nu_p)
    \Big)
    \\s.t.~ 
    &\theta_p\in(0,1)
\end{align*}

\noindent
where $\mathcal{B}(\cdot,~\cdot)$ is the beta function.
\\\\
For Bernoulli distribution and beta distribution, we have $\mathbb{E}[\mathbf{\pi}_{p}|\mathbf{\theta}_{p}]=\theta_p$. The portfolio switching strategy can be conceptualized as maximizing the projection from the raw blending ratio vector $\psi$ onto the parameter estimates vector $\hat{\theta}$:

\begin{align*}
    \hat{\psi}=~
    &\arg\max~
    \psi\cdot\hat{\theta}
    \\s.t.~ 
    & \| \psi \|_1=1, \psi_p>0
    \\
    \hat{\psi}_p=~
    &\mathbb{I}\{
    p=\arg\max_q~\hat{\theta}_q
    % \hat{\theta}_p=\max_q{\hat{\theta}_q}
    \}
\end{align*}

\subsection{Backtest}
\begin{figure}
    \centering
    \includegraphics[width=0.8\textwidth]{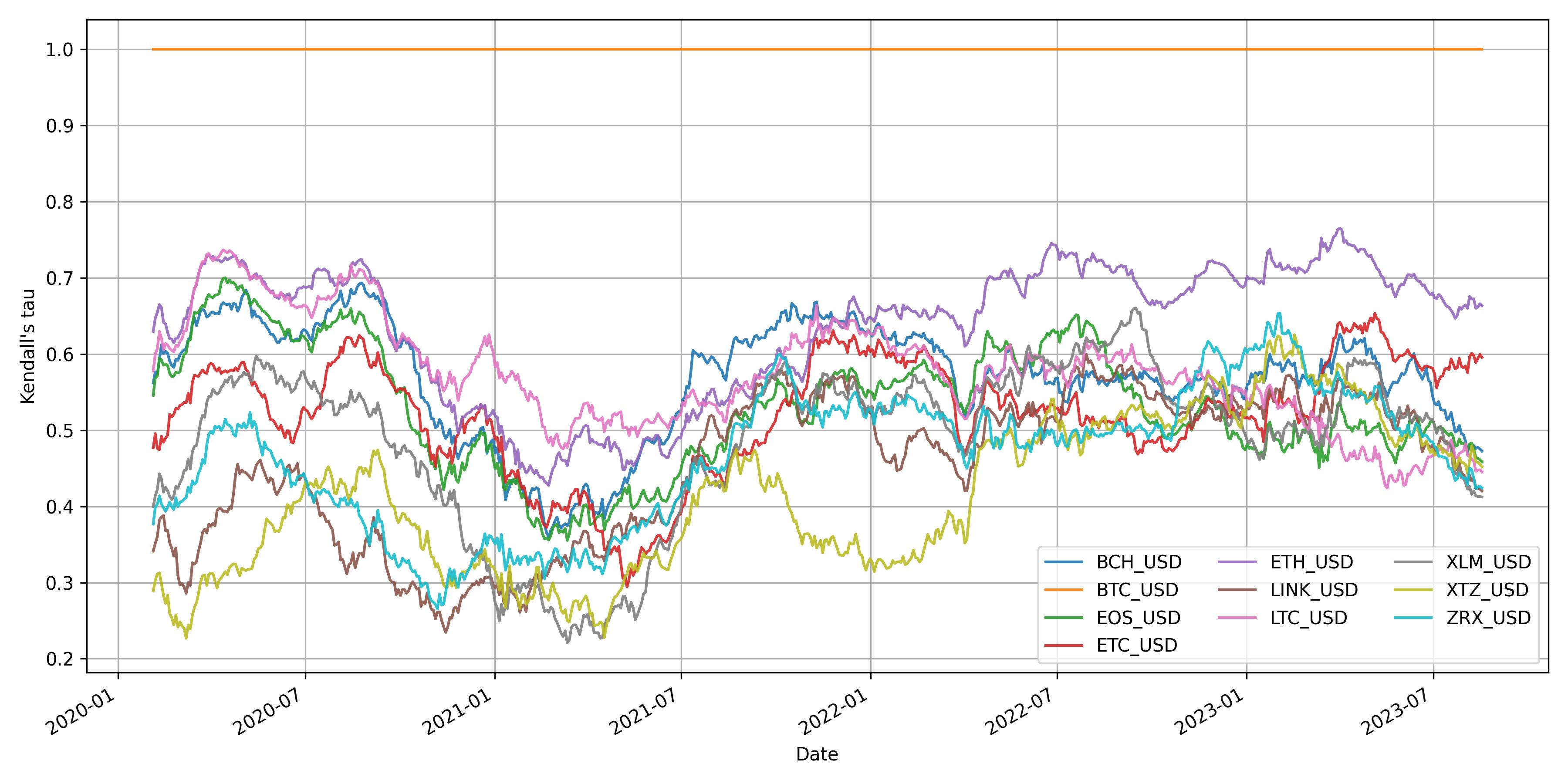}
    \caption{Kendall's $\tau$ between BTC and studied crypto-currencies, using a $182$-day rolling window of $2$-day simple returns as priced in USD on Coinbase.}
    \label{fig:blend_tau}
\end{figure}

\noindent
To backtest eclectic portfolios, we retrieved price time series from Coinbase (we use a different exchange for this backtest, just to be sure our results are not sensitive to the data sources), spanning from 2019-08-08 to 2023-08-17, covering $10$ currencies priced in USD including BCH, BTC, EOS, ETC, ETH, LINK, LTC, XLM, XTZ, and ZRX. As depicted in \cref{fig:blend_tau}, asset returns have notable high correlations. Rebalancing occurs every two days. At every step, we solve optimal portfolio weights $\mathbf{w}_{0,p}$ of each arm according to different generative model forecasts and proxy objective functions. We then calculate optimal blending ratios $\psi$ according to different similarity functions $s_{p}$, activation functions $\pi_p$, decay factors $\gamma\in\{0.9, 0.99, 0.999\}$, and blending methods (blending or switching). The rolling window length of historical asset returns for generative model fitting is $91$ steps or $182$ days. The rolling window length of historical $\pi_p$ for $\theta_p$ estimation and $\psi$ calculation is $26$ steps or $52$ days. The transaction cost $c$ is set as $50$ bps, together with the number of assets $D=10$ and the constraint boundary multiplier $m=5$.
To ensure robustness and assess performance consistency, we conduct multiple independent simulations using different random seeds In total, we have $5460$ paths and each has $619$ steps.

\begin{figure}
    \centering
    \includegraphics[width=0.8\textwidth]{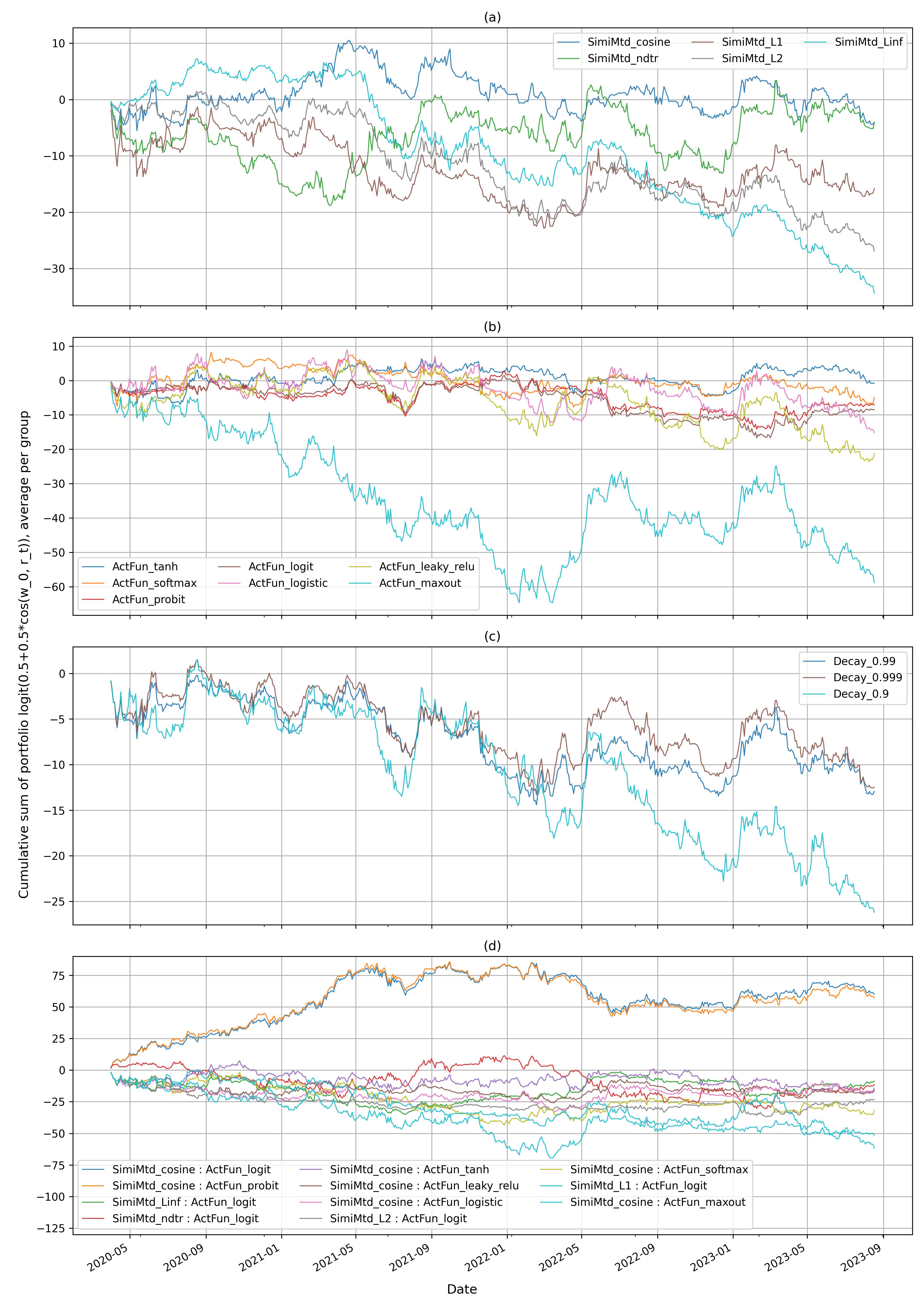}
    \caption{
    Cumulative sum of logit-cosine performance measure, averaged for eclectic portfolios as grouped by (a) similarity, (b) activation function, (c) decay factor, and (d) combination of similarity and activation function that use either cosine similarity or logit activation function.
    }
    \label{fig:blend_cumsum_logitcos}
\end{figure}

\noindent\\
For performance attribution, we fit \gls{lasso} (\ref{eq:lasso}) to eclectic portfolio performances per step, similar to what we have done for evaluating diverse pairings of generative model forecasts and objective functions
used for portfolio optimization in the previous section. We use the eclectic portfolio simple return $r_p$ (\ref{eq:rp}) and the logit of cosine similarity between eclectic portfolio weights and asset returns (\ref{eq:logit_cosine}) as the performance measure $y$. The design matrix ${\mathbf{X}} \in \mathbb{R}^{3379740\times 119}$ is filled with binary values ($0$ or $1$) encoding the presence of the foundation elements used in the eclectic portfolio construction: intercept, various similarity functions, optimality activation functions, decay factors, blending methods, and their two-variable interaction terms.
While $\mathbf{X}$ has $119$ columns, each row contains only eleven instances of the value $1$. The performance measures, represented as the column vector $\mathbf{y} \in \mathbb{R}^{3379740\times 1}$ encompass data from all steps in all simulation paths. The optimal $\lambda^*$ is selected based on $7$-fold \gls{cv} that minimizes \gls{mse}.
\\\\
We present the non-zero coefficients obtained at the optimal $\lambda^*$ in \gls{lasso} models for the $r_p$ in \cref{tab:blend_lasso_coef_rp} and for the logit-cosine performance measure in \cref{tab:blend_lasso_coef_logit_cosine}. 
The significance of interaction terms between similarities and activation functions are apparent with their positive coefficients persisting in \gls{lasso} model. The coefficients for the value models are of larger magnitude compared to those for the policy models; this indicates a good value model outweighs action selection policies. The specific value model combination of using cosine for similarity and logit activation function for optimality has the biggest coefficient in \gls{lasso} model for performance attribution. This combination also leads to the best performing eclectic portfolio as shown in \cref{fig:blend_cumsum_logitcos}d. However, the group average using logit activation function alone, does not stand out as shown in \cref{fig:blend_cumsum_logitcos}b. In general, most group-average logit-cosine performance measures exhibit negative trends. Furthermore, in \cref{fig:blend_cumsum_logitcos}c, we see that portfolios with a decay factor $\gamma=0.999$ showed better performances than those with smaller $\gamma$. A larger $\gamma$ lets more historical $\pi_p$ records effectively engage in the \gls{wmle} of $\hat{\theta}_p$ and the policy model of $\psi$. Though a shorter lookback window may lead to a swifter adaptation of the policy model, for this market a larger sample size is beneficial.
%%likely due to the low signal-to-noise ratio.

\begin{table}[!ht]
\centering\scriptsize
\caption{
Non-zero coefficients in \gls{lasso} for eclectic portfolio simple return, corresponding to combinations of similarities, activation functions, decay factors, and blending methods.
}
\label{tab:blend_lasso_coef_rp}
\begin{tabular}{lr}
\toprule
coefficient & value \\
\midrule
intercept & 0.052420 \\
SimiMtd cosine : ActFun logit & 0.040435 \\
SimiMtd L1 : ActFun probit & 0.032191 \\
SimiMtd ndtr : ActFun softmax & 0.029074 \\
SimiMtd ndtr : ActFun logistic & 0.021552 \\
SimiMtd ndtr : ActFun tanh & 0.021014 \\
ActFun logistic & 0.018417 \\
ActFun leaky relu & 0.016727 \\
SimiMtd Linf : ActFun maxout & 0.015468 \\
SimiMtd L2 : ActFun leaky relu & 0.013818 \\
SimiMtd cosine : ActFun logit & 0.012790 \\
SimiMtd Linf : ActFun leaky relu & 0.012582 \\
ActFun logit & 0.011888 \\
SimiMtd L2 : ActFun tanh & 0.008916 \\
SimiMtd L2 & 0.008098 \\
ActFun maxout & 0.005346 \\
SimiMtd L1 : ActFun maxout & 0.002466 \\
SimiMtd cosine : ActFun logistic & 0.001398 \\
SimiMtd L2 : ActFun softmax & 0.001231 \\
SimiMtd Linf : ActFun tanh & 0.000469 \\
SimiMtd L1 & 0.000295 \\
Decay 0.999 & 0.000245 \\
BldMtd switch & 0.000034 \\
Decay 0.9 & -0.000085 \\
SimiMtd Linf & -0.001580 \\
SimiMtd cosine & -0.001636 \\
SimiMtd Linf : ActFun probit & -0.001769 \\
SimiMtd L2 : ActFun maxout & -0.006973 \\
SimiMtd L1 : ActFun tanh & -0.008966 \\
SimiMtd L1 : ActFun leaky relu & -0.014248 \\
ActFun probit & -0.016101 \\
SimiMtd L1 : ActFun softmax & -0.016573 \\
SimiMtd ndtr : ActFun probit & -0.017719 \\
SimiMtd L2 : ActFun logit & -0.028180 \\
SimiMtd Linf : ActFun logit & -0.028317 \\
SimiMtd ndtr : ActFun logit & -0.045055 \\
\bottomrule
\end{tabular}
\end{table}

\begin{figure}
    \centering
    \includegraphics[width=0.8\textwidth]{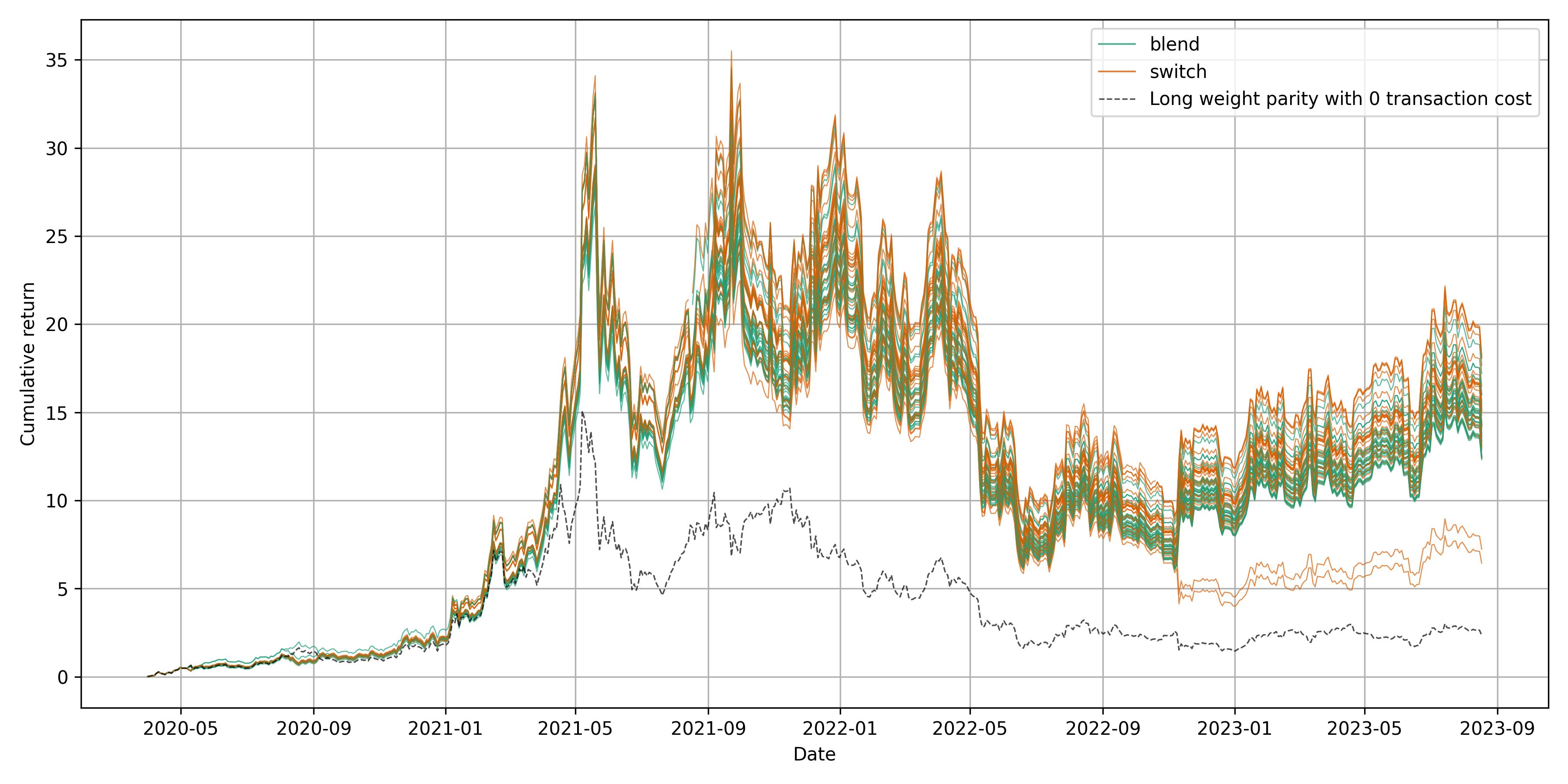}
    \caption{Visualization of eclectic portfolios using cosine similarity function and logit activation function with a decay factor $\gamma=0.999$, in cumulative simple returns.}
    \label{fig:blend_best}
\end{figure}

\begin{table}[!ht]
\centering\scriptsize
\caption{
Non-zero coefficients in \gls{lasso} for eclectic portfolio logit-cosine performance measure, corresponding to combinations of similarities, activation functions, decay factors, and blending methods.
}
\label{tab:blend_lasso_coef_logit_cosine}
\begin{tabular}{lr}
\toprule
coefficient & value \\
\midrule
SimiMtd cosine : ActFun logit & 0.694396 \\
SimiMtd cosine : ActFun probit & 0.683284 \\
SimiMtd ndtr : ActFun softmax & 0.629263 \\
SimiMtd L1 : ActFun logit & 0.583020 \\
SimiMtd L1 : ActFun probit & 0.574046 \\
ActFun logistic & 0.518490 \\
SimiMtd Linf : ActFun leaky relu & 0.387092 \\
ActFun leaky relu & 0.355478 \\
SimiMtd ndtr : ActFun tanh & 0.346860 \\
ActFun maxout & 0.276777 \\
SimiMtd L2 : ActFun tanh & 0.274927 \\
SimiMtd Linf : ActFun maxout & 0.264649 \\
intercept & 0.235726 \\
SimiMtd L2 : ActFun leaky relu & 0.183136 \\
SimiMtd L2 & 0.159240 \\
SimiMtd ndtr : ActFun logistic & 0.141152 \\
Decay 0.999 & 0.111536 \\
SimiMtd ndtr & 0.076147 \\
Decay 0.99 & 0.043022 \\
SimiMtd ndtr : ActFun maxout & 0.039835 \\
SimiMtd L2 : ActFun softmax & 0.004616 \\
BldMtd switch & 0.002090 \\
SimiMtd L1 : ActFun softmax & -0.016793 \\
SimiMtd cosine & -0.030179 \\
ActFun logit & -0.042316 \\
SimiMtd L2 : ActFun maxout & -0.112805 \\
ActFun softmax & -0.125102 \\
SimiMtd cosine : ActFun tanh & -0.176605 \\
SimiMtd Linf : ActFun softmax & -0.217533 \\
SimiMtd ndtr : ActFun leaky relu & -0.259776 \\
SimiMtd L2 : ActFun logit & -0.378502 \\
SimiMtd L2 : ActFun probit & -0.411934 \\
SimiMtd Linf : ActFun logit & -0.533289 \\
SimiMtd Linf : ActFun probit & -0.651316 \\
SimiMtd ndtr : ActFun logit & -1.109325 \\
SimiMtd ndtr : ActFun probit & -1.150317 \\
\bottomrule
\end{tabular}
\end{table}

\noindent\\
To provide a concrete example of the best-performing candidate identified through \gls{lasso} results, we present the backtest performances of portfolios using cosine for similarity and logit for optimality with a decay factor $\gamma=0.999$ in \cref{fig:blend_best}. It can be seen from the figure that all simulated paths consistently outperform the benchmark, which maintains long-weight parity but without including transaction costs in portfolio return. This figure also shows that the level of outperformance achieved by the eclectic portfolio to its benchmark significantly surpasses the outperformance achieved by any individual generative model-based portfolios to their benchmarks shown in \cref{fig:fix_best}. 

\begin{figure}
    \centering
    \includegraphics[width=.8\textwidth]{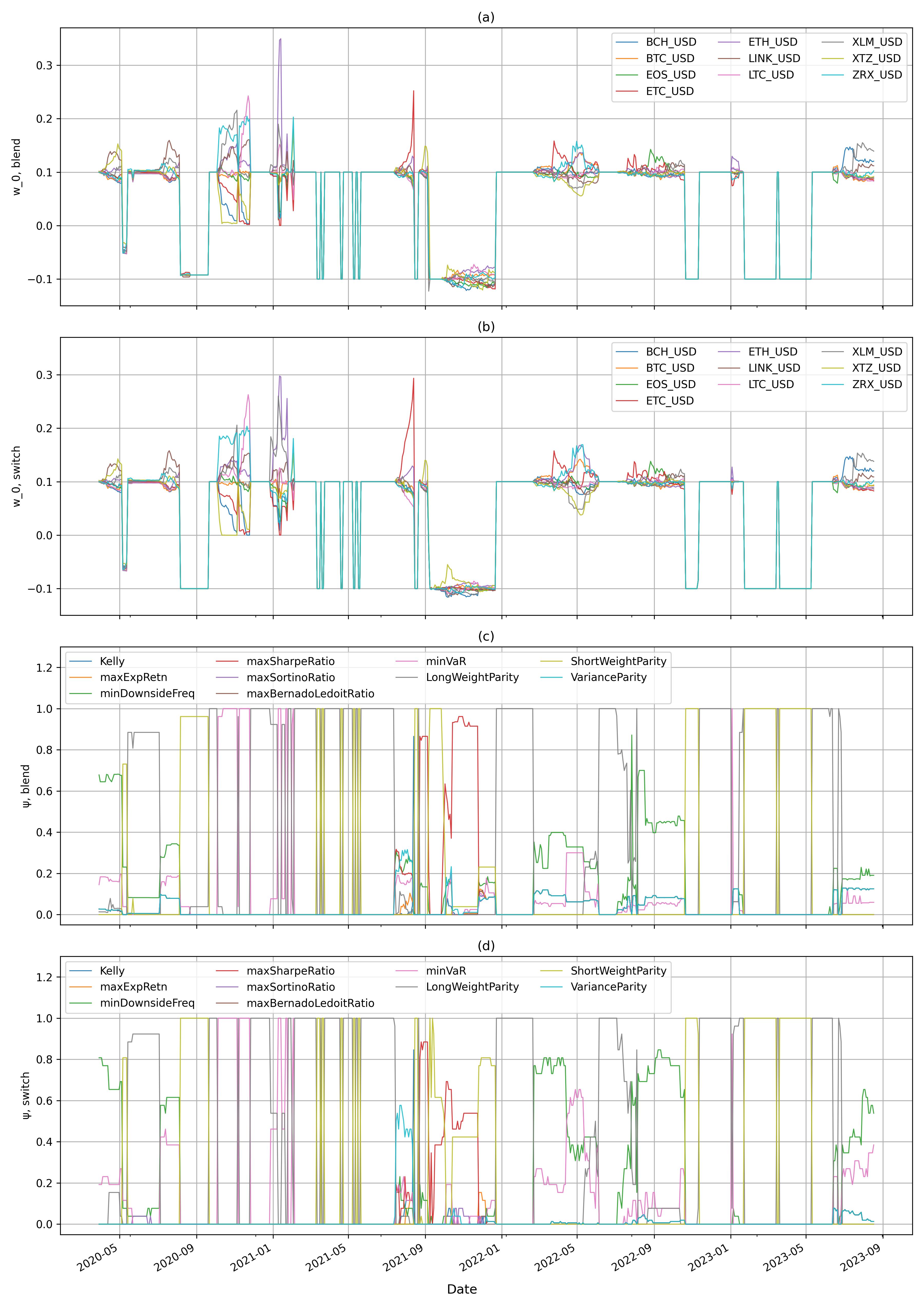}
    \caption{(a) Asset weights $\mathbf{w}_0$ for blending portfolios, (b) asset weights $\mathbf{w}_0$ for switching portfolios, (c) blending ratio $\psi$ for blending portfolios, and (d) blending ratio $\psi$ for switching portfolios. All in group average from eclectic portfolios using cosine similarity function and logit activation function with a decay factor $\gamma=0.999$.
    }
    \label{fig:blend_w0_psi}
\end{figure}

\begin{figure}
    \centering
    \includegraphics[width=0.8\textwidth]{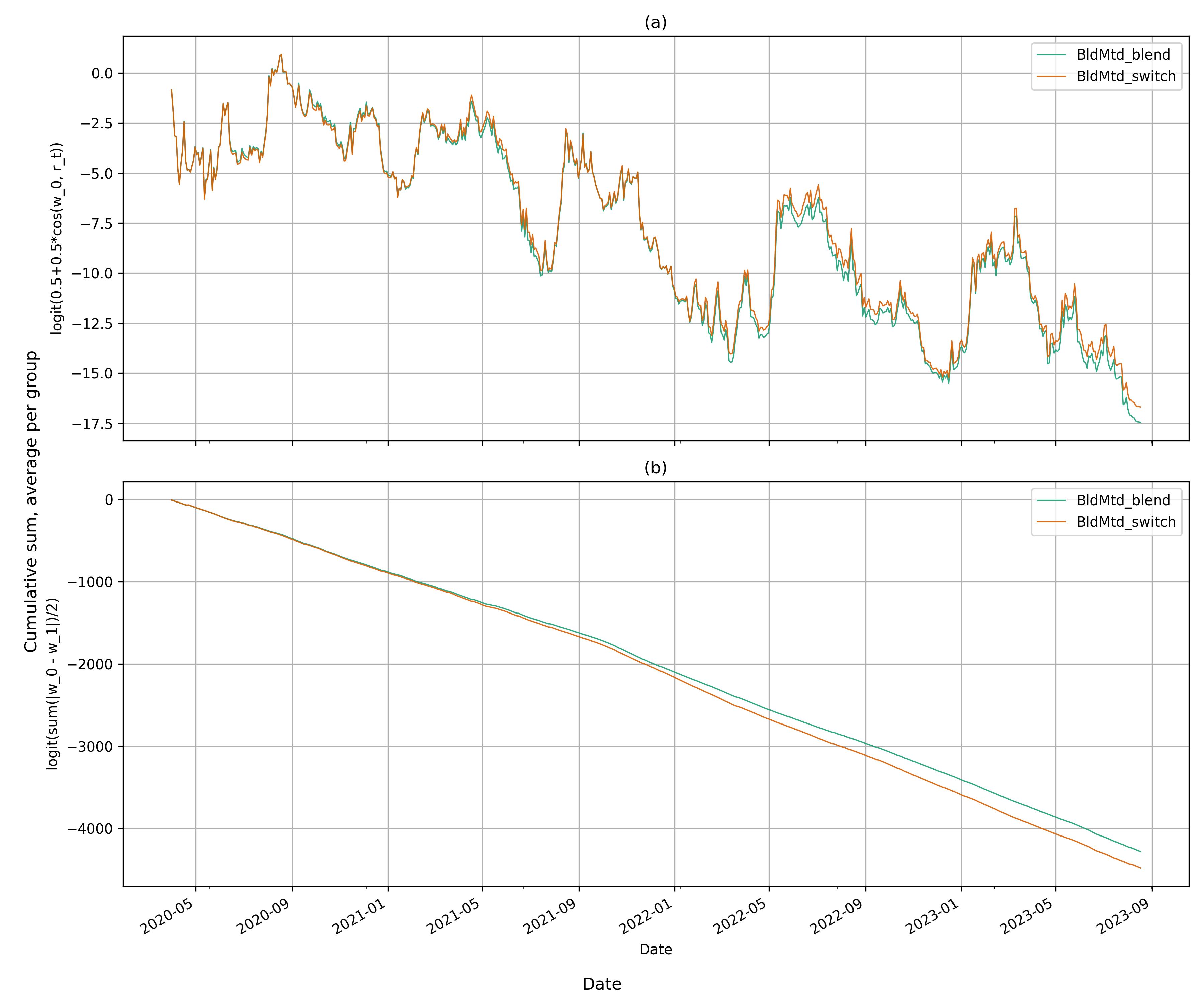}
    \caption{
    Cumulative sum of (a) logit-cosine performance measure and (b) logit-turnover, in group average from all eclectic portfolios.
    }
    \label{fig:blend_cumsum_bldmtd}
\end{figure}

\noindent\\
For portfolio groups using blending or switching, we chart the averaged final portfolio weights $\mathbf{w}_0$ and blending ratio $\psi$ in \cref{fig:blend_w0_psi}. We also plot their group averaged cumulative sum of the logit-cosine performance measure and the logit-turnover in \cref{fig:blend_cumsum_bldmtd}, where the logit-turnover \cref{eq:logit_turnover} is defined as:

\begin{align}
    \text{logit}
    \left(
    \frac{\|\mathbf{w}_{t,0}-\mathbf{w}_{t-1,1}\|_1}{2}
    \right)
    \label{eq:logit_turnover}
\end{align}

\noindent
As evident in \cref{fig:blend_w0_psi}, switching portfolios have more concentrated blending ratios $\psi$ than blending portfolios, and their asset weights $\mathbf{w}_0$ display a similar pattern. Both policy models allow for abrupt $\psi$ evolution and occasionally all-in short-weight parity portfolios, which is not common in traditional portfolio theory. It can be seen from \cref{fig:blend_cumsum_bldmtd}, switching portfolios have less turnover, higher $r_p$, and higher logit-cosine performance measure. 
%%Traditionally, Thompson sampling allocates $\psi$ to align the probability of each arm being optimal, while probability matching allocates $\psi$ to align the expected reward of each arm. In both cases, arms with small chances of being optimal or yielding little rewards may still be visited to update their posterior probability of being optimal. From another aspect, agents holding switching policy bear the risk of unintentionally concentrating on suboptimal arms due to inadequate value modeling, while agents holding blending policy bear the risk of intentionally diversifying on suboptimal arms due to inherent policy compromising. From the results above, with full feedback and a good value model, the agent is safe to use a switching policy and exploit concentration.

\section{Conclusion and future research}
\label{sec:conclusion}
\noindent
In this paper, we embark on a comprehensive exploration of our portfolio construction framework, delving into its foundation elements and conducting a thorough evaluation of its efficacy. Operating within the dynamic and evolving realm of cryptocurrencies, we craft a portfolio framework that is not only applicable to the cryptocurrency domain but also relevant to broader financial markets.
\\\\
Leveraging high-performance computing, we meticulously investigated diverse pairings of generative model forecasts and proxy objective functions. Our findings underscored the pivotal role played by proxy objective functions, as evidenced by their larger coefficient in performance \gls{lasso} models. Notably, the vine copula emerged as a superior choice for multivariate dependence modeling, and Sharpe ratio portfolios consistently outperformed alternative approaches.
\\\\
To align with a broad spectrum of investment strategies, we introduced eclectic portfolios, by extending the multi-armed bandit framework to portfolio blending. This approach blends optimal weights from different portfolios in a long-only manner, promoting diversification and leveraging historical feedback for refinement. We introduced similarity and optimality measures for value models and then employed probability-matching (``blending") and a greedy algorithm (``switching") for policy models. Notably, the specific value model using cosine similarity and logit activation function for optimality consistently delivered robust coefficients in the \gls{lasso} analysis. Switching portfolios exhibited higher variability but also achieved superior average performance compared to blending portfolios. The extent of outperformance by eclectic portfolios over their benchmarks significantly surpassed that achieved by individual generative model-based portfolios over their respective benchmarks.
\\\\
In summary, our exploration of generative model-based cryptocurrency portfolios illustrates the potential for portfolio diversification and optimization through our framework. The versatility of this approach, coupled with its capacity to leverage historical full feedback, positions eclectic portfolios as an attractive option for portfolio strategies.
\\\\
Looking ahead to future research, we envision the expansion of this framework to encompass various decision processes, including causal decision theory and Delphi methods \citep{dalkey1963experimental}. The framework can be extended to incorporate proprietary alpha models into returns matrix $R^\mathbb{P}$ for portfolio construction \citep{garthwaite_quantifying_1996}. Further investigation into proxy objective functions tailored to address tail risks with a Pareto distribution assumption \citep{chen_measuring_2022} could provide valuable insights into the widely recognized ``lollapalooza" effect. 
%%Refinement of value models using artificial neural networks holds promise for future exploration. 
The policy model, which relies on binomial, categorical, and beta distributions can benefit from conjugate priors and Bayesian learning. In addition, different investment mandates such as those centered on ESG or CSR are existing candidates to blend for ethical and sustainability objectives. 
%%In practice, firms have a more extensive pool of portfolio candidates to blend than explored in this study. Designing an effective topological sorting mechanism for the hierarchical and iterative policy model becomes imperative. 
Furthermore, addressing multi-step asset return forecasts and solving multi-step portfolio optimization (\ref{eq:objfun_ultimate}), potentially utilizing dynamic programming \citep{isichenko_quantitative_2021}, represents both an opportunity and a computational challenge. 
%%Streamlining the portfolio candidates and imposing reasonable constraints on $\mathbf{w}_0$ and $\psi$ will be crucial for expediting the backtesting and portfolio construction processes.

\section*{Acknowledgment}
\noindent
We thank Mr. Linghao Huang for his help in validating the \gls{dcc}-\gls{garch} model as well as Mr. Kaihao Chen and Mr. Xuxin Gao for their help in validating backtest processes. We appreciate the two anonymous reviewers for their constructive comments and inspiring questions.
%%This research is partly supported by a Ministry of Education academic research grant on Outbreak Resilience, Forecasting, and Early Warning.

% \\\\


% \bibliographystyle{elsarticle-num-names}
\bibliographystyle{elsarticle-harv}
\bibliography{1_references}

\begin{thebibliography}{62}
\expandafter\ifx\csname natexlab\endcsname\relax\def\natexlab#1{#1}\fi
\providecommand{\url}[1]{\texttt{#1}}
\providecommand{\href}[2]{#2}
\providecommand{\path}[1]{#1}
\providecommand{\DOIprefix}{doi:}
\providecommand{\ArXivprefix}{arXiv:}
\providecommand{\URLprefix}{URL: }
\providecommand{\Pubmedprefix}{pmid:}
\providecommand{\doi}[1]{\href{http://dx.doi.org/#1}{\path{#1}}}
\providecommand{\Pubmed}[1]{\href{pmid:#1}{\path{#1}}}
\providecommand{\bibinfo}[2]{#2}
\ifx\xfnm\relax \def\xfnm[#1]{\unskip,\space#1}\fi
%Type = Article
\bibitem[{Acerbi(2002)}]{acerbi_spectral_2002}
\bibinfo{author}{Acerbi, C.}, \bibinfo{year}{2002}.
\newblock \bibinfo{title}{Spectral measures of risk: {A} coherent
  representation of subjective risk aversion}.
\newblock \bibinfo{journal}{Journal of Banking \& Finance}
  \bibinfo{volume}{26}, \bibinfo{pages}{1505--1518}.
\newblock \DOIprefix\doi{https://doi.org/10.1016/S0378-4266(02)00281-9}.
%Type = Incollection
\bibitem[{Akaike(1998)}]{akaike_information_1998}
\bibinfo{author}{Akaike, H.}, \bibinfo{year}{1998}.
\newblock \bibinfo{title}{Information theory and an extension of the maximum
  likelihood principle}, in: \bibinfo{editor}{Parzen, E.},
  \bibinfo{editor}{Tanabe, K.}, \bibinfo{editor}{Kitagawa, G.} (Eds.),
  \bibinfo{booktitle}{Selected {Papers} of {Hirotugu} {Akaike}}.
  \bibinfo{publisher}{Springer New York}, \bibinfo{address}{New York, NY}, pp.
  \bibinfo{pages}{199--213}.
\newblock \DOIprefix\doi{10.1007/978-1-4612-1694-0_15}.
%Type = Article
\bibitem[{Bernardo and Ledoit(2000)}]{bernardo_gain_2000}
\bibinfo{author}{Bernardo, A.E.}, \bibinfo{author}{Ledoit, O.},
  \bibinfo{year}{2000}.
\newblock \bibinfo{title}{Gain, loss, and asset pricing}.
\newblock \bibinfo{journal}{Journal of Political Economy}
  \bibinfo{volume}{108}, \bibinfo{pages}{144--172}.
\newblock \DOIprefix\doi{10.1086/262114}.
%Type = Article
\bibitem[{Bollerslev(1986)}]{bollerslev_generalized_1986}
\bibinfo{author}{Bollerslev, T.}, \bibinfo{year}{1986}.
\newblock \bibinfo{title}{Generalized autoregressive conditional
  heteroskedasticity}.
\newblock \bibinfo{journal}{Journal of Econometrics} \bibinfo{volume}{31},
  \bibinfo{pages}{307--327}.
\newblock \DOIprefix\doi{https://doi.org/10.1016/0304-4076(86)90063-1}.
%Type = Article
\bibitem[{Chen and Cheng(2022)}]{chen_measuring_2022}
\bibinfo{author}{Chen, K.}, \bibinfo{author}{Cheng, T.}, \bibinfo{year}{2022}.
\newblock \bibinfo{title}{Measuring tail risks}.
\newblock \bibinfo{journal}{The Journal of Finance and Data Science}
  \bibinfo{volume}{8}, \bibinfo{pages}{296--308}.
\newblock \DOIprefix\doi{10.1016/j.jfds.2022.11.001}.
%Type = Article
\bibitem[{Czado(2019)}]{czado_analyzing_2019}
\bibinfo{author}{Czado, C.}, \bibinfo{year}{2019}.
\newblock \bibinfo{title}{Analyzing dependent data with vine copulas: {A}
  practical guide with {R}}.
\newblock \bibinfo{journal}{Lecture Notes in Statistics, Springer}
  \bibinfo{volume}{222}.
\newblock \DOIprefix\doi{https://doi.org/10.1007/978-3-030-13785-4}.
%Type = Article
\bibitem[{Czado et~al.(2022)Czado, Bax, Sahin, Nagler, Min and
  Paterlini}]{czado_vine_2022}
\bibinfo{author}{Czado, C.}, \bibinfo{author}{Bax, K.}, \bibinfo{author}{Sahin,
  {\"O}.}, \bibinfo{author}{Nagler, T.}, \bibinfo{author}{Min, A.},
  \bibinfo{author}{Paterlini, S.}, \bibinfo{year}{2022}.
\newblock \bibinfo{title}{Vine copula based dependence modeling in sustainable
  finance}.
\newblock \bibinfo{journal}{The Journal of Finance and Data Science}
  \bibinfo{volume}{8}, \bibinfo{pages}{309--330}.
\newblock \DOIprefix\doi{https://doi.org/10.1016/j.jfds.2022.11.003}.
%Type = Article
\bibitem[{Dalkey and Helmer(1963)}]{dalkey1963experimental}
\bibinfo{author}{Dalkey, N.}, \bibinfo{author}{Helmer, O.},
  \bibinfo{year}{1963}.
\newblock \bibinfo{title}{An experimental application of the delphi method to
  the use of experts}.
\newblock \bibinfo{journal}{Management science} \bibinfo{volume}{9},
  \bibinfo{pages}{458--467}.
%Type = Article
\bibitem[{De~Prado(2018)}]{de_prado_10_2018}
\bibinfo{author}{De~Prado, M.L.}, \bibinfo{year}{2018}.
\newblock \bibinfo{title}{The 10 reasons most machine learning funds fail}.
\newblock \bibinfo{journal}{The Journal of Portfolio Management}
  \bibinfo{volume}{44}, \bibinfo{pages}{120--133}.
\newblock \DOIprefix\doi{10.3905/jpm.2018.44.6.120}.
%Type = Article
\bibitem[{Dißmann et~al.(2013)Dißmann, Brechmann, Czado and
  Kurowicka}]{dismann_selecting_2013}
\bibinfo{author}{Dißmann, J.}, \bibinfo{author}{Brechmann, E.C.},
  \bibinfo{author}{Czado, C.}, \bibinfo{author}{Kurowicka, D.},
  \bibinfo{year}{2013}.
\newblock \bibinfo{title}{Selecting and estimating regular vine copulae and
  application to financial returns}.
\newblock \bibinfo{journal}{Computational Statistics \& Data Analysis}
  \bibinfo{volume}{59}, \bibinfo{pages}{52--69}.
\newblock \DOIprefix\doi{https://doi.org/10.1016/j.csda.2012.08.010}.
%Type = Article
\bibitem[{Efron et~al.(2004)Efron, Hastie, Johnstone and
  Tibshirani}]{efron_least_2004}
\bibinfo{author}{Efron, B.}, \bibinfo{author}{Hastie, T.},
  \bibinfo{author}{Johnstone, I.}, \bibinfo{author}{Tibshirani, R.},
  \bibinfo{year}{2004}.
\newblock \bibinfo{title}{Least angle regression}.
\newblock \bibinfo{journal}{The Annals of Statistics} \bibinfo{volume}{32},
  \bibinfo{pages}{407 -- 499}.
\newblock \DOIprefix\doi{10.1214/009053604000000067}.
%Type = Article
\bibitem[{Engle(2002)}]{engle_dynamic_2002}
\bibinfo{author}{Engle, R.}, \bibinfo{year}{2002}.
\newblock \bibinfo{title}{Dynamic conditional correlation: {A} simple class of
  multivariate generalized autoregressive conditional heteroskedasticity
  models}.
\newblock \bibinfo{journal}{Journal of Business \& Economic Statistics}
  \bibinfo{volume}{20}, \bibinfo{pages}{339--350}.
\newblock \URLprefix \url{http://www.jstor.org/stable/1392121}.
%Type = Article
\bibitem[{Fernholz(1999)}]{fernholz_diversity_1999}
\bibinfo{author}{Fernholz, R.}, \bibinfo{year}{1999}.
\newblock \bibinfo{title}{On the diversity of equity markets}.
\newblock \bibinfo{journal}{Journal of Mathematical Economics}
  \bibinfo{volume}{31}, \bibinfo{pages}{393--417}.
\newblock \DOIprefix\doi{https://doi.org/10.1016/S0304-4068(97)00018-9}.
%Type = Article
\bibitem[{Friedman and Savage(1948)}]{friedman_utility_1948}
\bibinfo{author}{Friedman, M.}, \bibinfo{author}{Savage, L.J.},
  \bibinfo{year}{1948}.
\newblock \bibinfo{title}{The utility analysis of choices involving risk}.
\newblock \bibinfo{journal}{Journal of Political Economy} \bibinfo{volume}{56},
  \bibinfo{pages}{279--304}.
\newblock \URLprefix \url{http://www.jstor.org/stable/1826045}.
%Type = Inproceedings
\bibitem[{Fujishima and Nakagawa(2022)}]{fujishima2022multiple}
\bibinfo{author}{Fujishima, K.}, \bibinfo{author}{Nakagawa, K.},
  \bibinfo{year}{2022}.
\newblock \bibinfo{title}{Multiple portfolio blending strategy with thompson
  sampling}, in: \bibinfo{booktitle}{2022 12th International Congress on
  Advanced Applied Informatics (IIAI-AAI)}, \bibinfo{organization}{IEEE}. pp.
  \bibinfo{pages}{449--454}.
%Type = Article
\bibitem[{Garthwaite and Dickey(1996)}]{garthwaite_quantifying_1996}
\bibinfo{author}{Garthwaite, P.H.}, \bibinfo{author}{Dickey, J.M.},
  \bibinfo{year}{1996}.
\newblock \bibinfo{title}{Quantifying and using expert opinion for
  variable-selection problems in regression}.
\newblock \bibinfo{journal}{Chemometrics and Intelligent Laboratory Systems}
  \bibinfo{volume}{35}, \bibinfo{pages}{1--26}.
\newblock \DOIprefix\doi{https://doi.org/10.1016/S0169-7439(96)00035-4}.
%Type = Article
\bibitem[{Heckerman and Shachter(1995)}]{heckerman_decision-theoretic_1995}
\bibinfo{author}{Heckerman, D.}, \bibinfo{author}{Shachter, R.},
  \bibinfo{year}{1995}.
\newblock \bibinfo{title}{Decision-theoretic foundations for causal reasoning}.
\newblock \bibinfo{journal}{J. Artif. Int. Res.} \bibinfo{volume}{3},
  \bibinfo{pages}{405--430}.
%Type = Article
\bibitem[{Howard et~al.(1972)Howard, Matheson and North}]{howard_decision_1972}
\bibinfo{author}{Howard, R.A.}, \bibinfo{author}{Matheson, J.E.},
  \bibinfo{author}{North, D.W.}, \bibinfo{year}{1972}.
\newblock \bibinfo{title}{The decision to seed hurricanes}.
\newblock \bibinfo{journal}{Science} \bibinfo{volume}{176},
  \bibinfo{pages}{1191--1202}.
\newblock \DOIprefix\doi{10.1126/science.176.4040.1191}.
%Type = Book
\bibitem[{Isichenko(2021)}]{isichenko_quantitative_2021}
\bibinfo{author}{Isichenko, M.}, \bibinfo{year}{2021}.
\newblock \bibinfo{title}{Quantitative Portfolio Management: The Art and
  Science of Statistical Arbitrage}.
\newblock \bibinfo{publisher}{Wiley}.
%Type = Book
\bibitem[{Joe(2014)}]{joe_dependence_2014}
\bibinfo{author}{Joe, H.}, \bibinfo{year}{2014}.
\newblock \bibinfo{title}{Dependence Modeling with Copulas}.
\newblock Chapman \& Hall/CRC Monographs on Statistics \& Applied Probability,
  \bibinfo{publisher}{Taylor \& Francis}.
%Type = Article
\bibitem[{Kahneman and Tversky(1979)}]{kahneman_prospect_1979}
\bibinfo{author}{Kahneman, D.}, \bibinfo{author}{Tversky, A.},
  \bibinfo{year}{1979}.
\newblock \bibinfo{title}{Prospect {Theory}: {An} {Analysis} of {Decision}
  under {Risk}}.
\newblock \bibinfo{journal}{Econometrica} \bibinfo{volume}{47},
  \bibinfo{pages}{263--291}.
\newblock \URLprefix \url{http://www.jstor.org/stable/1914185}.
%Type = Incollection
\bibitem[{Karatzas and Fernholz(2009)}]{bensoussan_stochastic_2009}
\bibinfo{author}{Karatzas, I.}, \bibinfo{author}{Fernholz, R.},
  \bibinfo{year}{2009}.
\newblock \bibinfo{title}{Stochastic portfolio theory: {An} overview}, in:
  \bibinfo{editor}{Bensoussan, A.}, \bibinfo{editor}{Zhang, Q.} (Eds.),
  \bibinfo{booktitle}{Special {Volume}: {Mathematical} {Modeling} and
  {Numerical} {Methods} in {Finance}}. \bibinfo{publisher}{Elsevier}.
  volume~\bibinfo{volume}{15} of \textit{\bibinfo{series}{Handbook of
  {Numerical} {Analysis}}}, pp. \bibinfo{pages}{89--167}.
\newblock \DOIprefix\doi{https://doi.org/10.1016/S1570-8659(08)00003-3}.
%Type = Article
\bibitem[{Kelly~Jr.(1956)}]{kelly_new_1956}
\bibinfo{author}{Kelly~Jr., J.L.}, \bibinfo{year}{1956}.
\newblock \bibinfo{title}{A new interpretation of information rate}.
\newblock \bibinfo{journal}{Bell System Technical Journal}
  \bibinfo{volume}{35}, \bibinfo{pages}{917--926}.
\newblock \DOIprefix\doi{https://doi.org/10.1002/j.1538-7305.1956.tb03809.x}.
%Type = Article
\bibitem[{Kolm et~al.(2021)Kolm, Ritter and
  Simonian}]{kolm_blacklitterman_2021}
\bibinfo{author}{Kolm, P.N.}, \bibinfo{author}{Ritter, G.},
  \bibinfo{author}{Simonian, J.}, \bibinfo{year}{2021}.
\newblock \bibinfo{title}{Black–litterman and beyond: {The} bayesian paradigm
  in investment management}.
\newblock \bibinfo{journal}{The Journal of Portfolio Management}
  \bibinfo{volume}{47}, \bibinfo{pages}{91--113}.
%Type = Article
\bibitem[{Kolm et~al.(2014)Kolm, Tütüncü and Fabozzi}]{kolm_60_2014}
\bibinfo{author}{Kolm, P.N.}, \bibinfo{author}{Tütüncü, R.},
  \bibinfo{author}{Fabozzi, F.J.}, \bibinfo{year}{2014}.
\newblock \bibinfo{title}{60 years of portfolio optimization: Practical
  challenges and current trends}.
\newblock \bibinfo{journal}{European Journal of Operational Research}
  \bibinfo{volume}{234}, \bibinfo{pages}{356--371}.
\newblock \DOIprefix\doi{https://doi.org/10.1016/j.ejor.2013.10.060}.
%Type = Article
\bibitem[{Lewis(1981)}]{lewis_causal_1981}
\bibinfo{author}{Lewis, D.}, \bibinfo{year}{1981}.
\newblock \bibinfo{title}{Causal decision theory}.
\newblock \bibinfo{journal}{Australasian Journal of Philosophy}
  \bibinfo{volume}{59}, \bibinfo{pages}{5--30}.
\newblock \DOIprefix\doi{10.1080/00048408112340011}.
%Type = Article
\bibitem[{Lezmi et~al.(2022)Lezmi, Roncalli and Xu}]{lezmi_multi-period_2022}
\bibinfo{author}{Lezmi, E.}, \bibinfo{author}{Roncalli, T.},
  \bibinfo{author}{Xu, J.}, \bibinfo{year}{2022}.
\newblock \bibinfo{title}{Multi-period portfolio optimization}.
\newblock \bibinfo{journal}{Available at SSRN} .
%Type = Book
\bibitem[{Lo(2017)}]{lo_adaptive_2017}
\bibinfo{author}{Lo, A.W.}, \bibinfo{year}{2017}.
\newblock \bibinfo{title}{Adaptive markets: Financial evolution at the speed of
  thought}.
\newblock \bibinfo{publisher}{Princeton University Press}.
\newblock \URLprefix \url{http://www.jstor.org/stable/j.ctvc7778k}.
%Type = Book
\bibitem[{Lo and Foerster(2021)}]{lo_pursuit_2021}
\bibinfo{author}{Lo, A.W.}, \bibinfo{author}{Foerster, S.R.},
  \bibinfo{year}{2021}.
\newblock \bibinfo{title}{In pursuit of the perfect portfolio}.
\newblock \bibinfo{publisher}{Princeton University Press},
  \bibinfo{address}{Princeton}.
\newblock \DOIprefix\doi{https://doi.org/10.1515/9780691222684}.
%Type = Article
\bibitem[{Lo et~al.(2021)Lo, Marlowe and Zhang}]{lo_maximize_2021}
\bibinfo{author}{Lo, A.W.}, \bibinfo{author}{Marlowe, K.P.},
  \bibinfo{author}{Zhang, R.}, \bibinfo{year}{2021}.
\newblock \bibinfo{title}{To maximize or randomize? an experimental study of
  probability matching in financial decision making}.
\newblock \bibinfo{journal}{PLOS ONE} \bibinfo{volume}{16},
  \bibinfo{pages}{1--20}.
\newblock \DOIprefix\doi{10.1371/journal.pone.0252540}.
%Type = Article
\bibitem[{Longerstaey and
  Spencer(1996)}]{longerstaey_riskmetricstmtechnical_1996}
\bibinfo{author}{Longerstaey, J.}, \bibinfo{author}{Spencer, M.},
  \bibinfo{year}{1996}.
\newblock \bibinfo{title}{Riskmetrics—technical document}.
\newblock \bibinfo{journal}{Morgan Guaranty Trust Company of New York: New
  York} \bibinfo{volume}{51}, \bibinfo{pages}{54}.
%Type = Article
\bibitem[{Malevergne and Sornette(2005)}]{malevergne_higher-moment_2005}
\bibinfo{author}{Malevergne, Y.}, \bibinfo{author}{Sornette, D.},
  \bibinfo{year}{2005}.
\newblock \bibinfo{title}{Higher-moment portfolio theory}.
\newblock \bibinfo{journal}{The Journal of Portfolio Management}
  \bibinfo{volume}{31}, \bibinfo{pages}{49--55}.
\newblock \DOIprefix\doi{https://doi.org/10.3905/jpm.2005.570150}.
%Type = Article
\bibitem[{Markowitz(1952)}]{markowitz_portfolio_1952}
\bibinfo{author}{Markowitz, H.}, \bibinfo{year}{1952}.
\newblock \bibinfo{title}{Portfolio selection}.
\newblock \bibinfo{journal}{The Journal of Finance} \bibinfo{volume}{7},
  \bibinfo{pages}{77--91}.
\newblock \DOIprefix\doi{https://doi.org/10.1111/j.1540-6261.1952.tb01525.x}.
%Type = Article
\bibitem[{Markowitz(2006)}]{markowitz_finetti_2006}
\bibinfo{author}{Markowitz, H.}, \bibinfo{year}{2006}.
\newblock \bibinfo{title}{de {Finetti} scoops {Markowitz}}.
\newblock \bibinfo{journal}{Journal of Investment Management}
  \bibinfo{volume}{4}.
%Type = Article
\bibitem[{Markowitz(2014)}]{markowitz_meanvariance_2014}
\bibinfo{author}{Markowitz, H.}, \bibinfo{year}{2014}.
\newblock \bibinfo{title}{Mean–variance approximations to expected utility}.
\newblock \bibinfo{journal}{European Journal of Operational Research}
  \bibinfo{volume}{234}, \bibinfo{pages}{346--355}.
\newblock \DOIprefix\doi{https://doi.org/10.1016/j.ejor.2012.08.023}.
%Type = Article
\bibitem[{Markowitz(1999)}]{markowitz_early_1999}
\bibinfo{author}{Markowitz, H.M.}, \bibinfo{year}{1999}.
\newblock \bibinfo{title}{The early history of portfolio theory: 1600-1960}.
\newblock \bibinfo{journal}{Financial Analysts Journal} \bibinfo{volume}{55},
  \bibinfo{pages}{5--16}.
\newblock \URLprefix \url{http://www.jstor.org/stable/4480178}.
%Type = Article
\bibitem[{Markowitz(2010)}]{markowitz_portfolio_2010}
\bibinfo{author}{Markowitz, H.M.}, \bibinfo{year}{2010}.
\newblock \bibinfo{title}{Portfolio theory: {As} {I} still see it}.
\newblock \bibinfo{journal}{Annu. Rev. Financ. Econ.} \bibinfo{volume}{2},
  \bibinfo{pages}{1--23}.
%Type = Book
\bibitem[{McElreath(2020)}]{mcelreath_statistical_2020}
\bibinfo{author}{McElreath, R.}, \bibinfo{year}{2020}.
\newblock \bibinfo{title}{Statistical rethinking: {A} {Bayesian} course with
  examples in {R} and {Stan}}.
\newblock \bibinfo{publisher}{CRC press}.
%Type = Book
\bibitem[{Munger and Kaufman(2008)}]{munger_poor_2008}
\bibinfo{author}{Munger, C.T.}, \bibinfo{author}{Kaufman, P.},
  \bibinfo{year}{2008}.
\newblock \bibinfo{title}{Poor {Charlie}’s Almanack}.
\newblock \bibinfo{publisher}{Donning}.
%Type = Book
\bibitem[{von Neumann et~al.(1944)von Neumann, Morgenstern and
  Rubinstein}]{von_neumann_theory_1944}
\bibinfo{author}{von Neumann, J.}, \bibinfo{author}{Morgenstern, O.},
  \bibinfo{author}{Rubinstein, A.}, \bibinfo{year}{1944}.
\newblock \bibinfo{title}{Theory of Games and Economic Behavior (60th
  Anniversary Commemorative Edition)}.
\newblock \bibinfo{publisher}{Princeton University Press}.
\newblock \URLprefix \url{http://www.jstor.org/stable/j.ctt1r2gkx}.
%Type = Article
\bibitem[{Orskaug(2009)}]{orskaug_dcc-garch_2009}
\bibinfo{author}{Orskaug, E.}, \bibinfo{year}{2009}.
\newblock \bibinfo{title}{{DCC}-{GARCH} model-with various error
  distributions}.
\newblock \bibinfo{journal}{Norwegian Computing Center, Pub. no. SAMBA/19/09,
  Jun} .
%Type = Inproceedings
\bibitem[{Paolella and Polak(2018)}]{paolella_cobra_2018}
\bibinfo{author}{Paolella, M.S.}, \bibinfo{author}{Polak, P.},
  \bibinfo{year}{2018}.
\newblock \bibinfo{title}{Cobra: Copula-based portfolio optimization}, in:
  \bibinfo{editor}{Kreinovich, V.}, \bibinfo{editor}{Sriboonchitta, S.},
  \bibinfo{editor}{Chakpitak, N.} (Eds.), \bibinfo{booktitle}{Predictive
  econometrics and big data}, \bibinfo{publisher}{Springer International
  Publishing}, \bibinfo{address}{Cham}. pp. \bibinfo{pages}{36--77}.
%Type = Article
\bibitem[{Pearl(1995)}]{pearl_causal_1995}
\bibinfo{author}{Pearl, J.}, \bibinfo{year}{1995}.
\newblock \bibinfo{title}{{Causal diagrams for empirical research}}.
\newblock \bibinfo{journal}{Biometrika} \bibinfo{volume}{82},
  \bibinfo{pages}{669--688}.
\newblock \DOIprefix\doi{10.1093/biomet/82.4.669}.
%Type = Article
\bibitem[{Pedregosa et~al.(2011)Pedregosa, Varoquaux, Gramfort, Michel,
  Thirion, Grisel, Blondel, Prettenhofer, Weiss, Dubourg, Vanderplas, Passos,
  Cournapeau, Brucher, Perrot and Duchesnay}]{pedregosa_scikit-learn_2011}
\bibinfo{author}{Pedregosa, F.}, \bibinfo{author}{Varoquaux, G.},
  \bibinfo{author}{Gramfort, A.}, \bibinfo{author}{Michel, V.},
  \bibinfo{author}{Thirion, B.}, \bibinfo{author}{Grisel, O.},
  \bibinfo{author}{Blondel, M.}, \bibinfo{author}{Prettenhofer, P.},
  \bibinfo{author}{Weiss, R.}, \bibinfo{author}{Dubourg, V.},
  \bibinfo{author}{Vanderplas, J.}, \bibinfo{author}{Passos, A.},
  \bibinfo{author}{Cournapeau, D.}, \bibinfo{author}{Brucher, M.},
  \bibinfo{author}{Perrot, M.}, \bibinfo{author}{Duchesnay, E.},
  \bibinfo{year}{2011}.
\newblock \bibinfo{title}{Scikit-learn: Machine learning in python}.
\newblock \bibinfo{journal}{J. Mach. Learn. Res.} \bibinfo{volume}{12},
  \bibinfo{pages}{2825–2830}.
%Type = Article
\bibitem[{Qian(2005)}]{qian_financial_2005}
\bibinfo{author}{Qian, E.}, \bibinfo{year}{2005}.
\newblock \bibinfo{title}{On the financial interpretation of risk contribution:
  {Risk} budgets do add up}.
\newblock \bibinfo{journal}{Available at SSRN 684221} .
%Type = Article
\bibitem[{Qian(2011)}]{qian_risk_2011}
\bibinfo{author}{Qian, E.}, \bibinfo{year}{2011}.
\newblock \bibinfo{title}{Risk parity and diversification}.
\newblock \bibinfo{journal}{The Journal of Investing} \bibinfo{volume}{20},
  \bibinfo{pages}{119--127}.
%Type = Book
\bibitem[{Ramsey(1931)}]{ramsey_foundations_1931}
\bibinfo{author}{Ramsey, F.P.}, \bibinfo{year}{1931}.
\newblock \bibinfo{title}{The foundations of mathematical and other logical
  essays}.
\newblock \bibinfo{publisher}{Routledge and K. Paul}.
%Type = Article
\bibitem[{Rockafellar et~al.(2000)Rockafellar, Uryasev and
  {others}}]{rockafellar_optimization_2000}
\bibinfo{author}{Rockafellar, R.T.}, \bibinfo{author}{Uryasev, S.},
  \bibinfo{author}{{others}}, \bibinfo{year}{2000}.
\newblock \bibinfo{title}{Optimization of conditional value-at-risk}.
\newblock \bibinfo{journal}{Journal of risk} \bibinfo{volume}{2},
  \bibinfo{pages}{21--42}.
%Type = Article
\bibitem[{Rosenblatt(1952)}]{rosenblatt_remarks_1952}
\bibinfo{author}{Rosenblatt, M.}, \bibinfo{year}{1952}.
\newblock \bibinfo{title}{Remarks on a multivariate transformation}.
\newblock \bibinfo{journal}{The annals of mathematical statistics}
  \bibinfo{volume}{23}, \bibinfo{pages}{470--472}.
%Type = Article
\bibitem[{Roy(1952)}]{roy_safety_1952}
\bibinfo{author}{Roy, A.D.}, \bibinfo{year}{1952}.
\newblock \bibinfo{title}{Safety first and the holding of assets}.
\newblock \bibinfo{journal}{Econometrica: Journal of the econometric society} ,
  \bibinfo{pages}{431--449}.
%Type = Article
\bibitem[{Rubinstein(2002)}]{rubinstein_markowitzs_2002}
\bibinfo{author}{Rubinstein, M.}, \bibinfo{year}{2002}.
\newblock \bibinfo{title}{Markowitz's "portfolio selection": {A} fifty-year
  retrospective}.
\newblock \bibinfo{journal}{The Journal of finance} \bibinfo{volume}{57},
  \bibinfo{pages}{1041--1045}.
%Type = Article
\bibitem[{Savage(1951)}]{savage_theory_1951}
\bibinfo{author}{Savage, L.J.}, \bibinfo{year}{1951}.
\newblock \bibinfo{title}{The theory of statistical decision}.
\newblock \bibinfo{journal}{Journal of the American Statistical Association}
  \bibinfo{volume}{46}, \bibinfo{pages}{55--67}.
\newblock \DOIprefix\doi{10.1080/01621459.1951.10500768}.
%Type = Incollection
\bibitem[{Schmidt(2004)}]{schmidt_alternatives_2004}
\bibinfo{author}{Schmidt, U.}, \bibinfo{year}{2004}.
\newblock \bibinfo{title}{Alternatives to expected utility: Formal theories},
  in: \bibinfo{editor}{Barber{\`a}, S.}, \bibinfo{editor}{Hammond, P.J.},
  \bibinfo{editor}{Seidl, C.} (Eds.), \bibinfo{booktitle}{Handbook of Utility
  Theory: Volume 2 Extensions}. \bibinfo{publisher}{Springer US},
  \bibinfo{address}{Boston, MA}, pp. \bibinfo{pages}{757--837}.
\newblock \DOIprefix\doi{10.1007/978-1-4020-7964-1_2}.
%Type = Article
\bibitem[{Sharpe(1966)}]{sharpe_mutual_1966}
\bibinfo{author}{Sharpe, W.F.}, \bibinfo{year}{1966}.
\newblock \bibinfo{title}{Mutual fund performance}.
\newblock \bibinfo{journal}{The Journal of Business} \bibinfo{volume}{39},
  \bibinfo{pages}{119--138}.
\newblock \URLprefix \url{http://www.jstor.org/stable/2351741}.
%Type = Inproceedings
\bibitem[{Shen and Wang(2016)}]{shen_portfolio_2016}
\bibinfo{author}{Shen, W.}, \bibinfo{author}{Wang, J.}, \bibinfo{year}{2016}.
\newblock \bibinfo{title}{Portfolio blending via thompson sampling}, in:
  \bibinfo{booktitle}{Proceedings of the Twenty-Fifth International Joint
  Conference on Artificial Intelligence}, \bibinfo{publisher}{AAAI Press}. p.
  \bibinfo{pages}{1983–1989}.
%Type = Article
\bibitem[{Soros(2013)}]{soros_fallibility_2013}
\bibinfo{author}{Soros, G.}, \bibinfo{year}{2013}.
\newblock \bibinfo{title}{Fallibility, reflexivity, and the human uncertainty
  principle}.
\newblock \bibinfo{journal}{Journal of Economic Methodology}
  \bibinfo{volume}{20}, \bibinfo{pages}{309--329}.
%Type = Article
\bibitem[{Sortino and Price(1994)}]{sortino_performance_1994}
\bibinfo{author}{Sortino, F.A.}, \bibinfo{author}{Price, L.N.},
  \bibinfo{year}{1994}.
\newblock \bibinfo{title}{Performance measurement in a downside risk
  framework}.
\newblock \bibinfo{journal}{the Journal of Investing} \bibinfo{volume}{3},
  \bibinfo{pages}{59--64}.
\newblock \DOIprefix\doi{https://doi.org/10.3905/joi.3.3.59}.
%Type = Book
\bibitem[{Sutton and Barto(2018)}]{sutton_reinforcement_2018}
\bibinfo{author}{Sutton, R.S.}, \bibinfo{author}{Barto, A.G.},
  \bibinfo{year}{2018}.
\newblock \bibinfo{title}{Reinforcement learning: {An} introduction}.
\newblock \bibinfo{publisher}{MIT Press}.
%Type = Article
\bibitem[{Thompson(1933)}]{thompson_likelihood_1933}
\bibinfo{author}{Thompson, W.R.}, \bibinfo{year}{1933}.
\newblock \bibinfo{title}{On the likelihood that one unknown probability
  exceeds another in view of the evidence of two samples}.
\newblock \bibinfo{journal}{Biometrika} \bibinfo{volume}{25},
  \bibinfo{pages}{285--294}.
\newblock \URLprefix \url{http://www.jstor.org/stable/2332286}.
%Type = Incollection
\bibitem[{Thorp(1975)}]{thorp_portfolio_1975}
\bibinfo{author}{Thorp, E.O.}, \bibinfo{year}{1975}.
\newblock \bibinfo{title}{Portfolio choice and the kelly criterion}, in:
  \bibinfo{editor}{ZIEMBA, W.}, \bibinfo{editor}{VICKSON, R.} (Eds.),
  \bibinfo{booktitle}{Stochastic Optimization Models in Finance}.
  \bibinfo{publisher}{Academic Press}, pp. \bibinfo{pages}{599--619}.
\newblock \DOIprefix\doi{https://doi.org/10.1016/B978-0-12-780850-5.50051-4}.
%Type = Incollection
\bibitem[{Weinzierl(2022)}]{weinzierl_pillars_2022}
\bibinfo{author}{Weinzierl, T.}, \bibinfo{year}{2022}.
\newblock \bibinfo{title}{The pillars of science}, in:
  \bibinfo{booktitle}{Principles of parallel scientific computing: A first
  guide to numerical concepts and programming methods}.
  \bibinfo{publisher}{Springer Nature}, pp. \bibinfo{pages}{3--9}.
%Type = Article
\bibitem[{Xidonas et~al.(2020)Xidonas, Steuer and
  Hassapis}]{xidonas_robust_2020}
\bibinfo{author}{Xidonas, P.}, \bibinfo{author}{Steuer, R.},
  \bibinfo{author}{Hassapis, C.}, \bibinfo{year}{2020}.
\newblock \bibinfo{title}{Robust portfolio optimization: a categorized
  bibliographic review}.
\newblock \bibinfo{journal}{Annals of Operations Research}
  \bibinfo{volume}{292}, \bibinfo{pages}{533--552}.
\newblock \DOIprefix\doi{10.1007/s10479-020-03630-8}.

\end{thebibliography}





% \end{thebibliography}
\end{document}